\documentclass[prl,superscriptaddress]{revtex4}

\usepackage[T1]{fontenc}

\usepackage{physics}
\usepackage{etoolbox}
\usepackage{orcidlink}

\usepackage{graphicx}
\usepackage{dcolumn}
\usepackage{array}
\usepackage{tabularx}

\usepackage{floatrow}

\usepackage{eucal}
\usepackage[dvips]{epsfig}
\usepackage{amssymb}
\usepackage{amsfonts}
\usepackage{multirow}
\usepackage{chngcntr}

\usepackage[parfill]{parskip}    

\usepackage{amsmath,amsthm}
\usepackage{cleveref}
\usepackage{siunitx}

\usepackage{comment}

\newcommand{\be}{\begin{eqnarray}}
\newcommand{\ee}{\end{eqnarray}}
\newcommand{\bse}{\begin{subequations}}
\newcommand{\ese}{\end{subequations}}

\newcommand{\bnum}{\begin{enumerate}}
\newcommand{\enum}{\end{enumerate}}

\newcommand{\bit}{\begin{itemize}}
\newcommand{\eit}{\end{itemize}}

\newcommand{\bc}{\begin{cases}}
\newcommand{\ec}{\end{cases}}

\newcommand{\bpm}{\begin{pmatrix}}
\newcommand{\epm}{\end{pmatrix}}

\newcommand{\bvm}{\begin{vmatrix}}
\newcommand{\evm}{\end{vmatrix}}

\newcommand{\tn}{\textnormal}

\newcommand{\SNR}{\mathrm{SNR}}

\usepackage{xcolor}

\DeclareSIUnit\chains{chains}
\DeclareSIUnit\filaments{filaments}

\begin{document}

\title{
Slower swimming promotes chemotactic encounters between bacteria and small phytoplankton
}

\author{Riccardo Foffi\,\orcidlink{0000-0001-9568-0480}}
\affiliation{Institute of Environmental Engineering, Department of Civil, Environmental and Geomatic Engineering, ETH Zurich, Zurich, Switzerland}
\author{Douglas R. Brumley\,\orcidlink{0000-0003-0587-0251}}
\affiliation{School of Mathematics and
Statistics, The University of Melbourne, Parkville, Victoria, Australia}
\author{François Peaudecerf\,\orcidlink{0000-0003-0295-4556}}
\affiliation{Univ Rennes, CNRS, IPR (Institut de Physique de Rennes) - UMR 6251, F-35000 Rennes, France}
\author{Roman Stocker\,\orcidlink{0000-0002-3199-0508}}
\affiliation{Institute of Environmental Engineering, Department of Civil, Environmental and Geomatic Engineering, ETH Zurich, Zurich, Switzerland}
\author{Jonasz S\l{}omka\,\orcidlink{0000-0002-7097-5810}}
\affiliation{Institute of Environmental Engineering, Department of Civil, Environmental and Geomatic Engineering, ETH Zurich, Zurich, Switzerland}

\date{\today}
\begin{abstract}
Chemotaxis enables marine bacteria to increase encounters with phytoplankton cells by reducing their search times, provided that bacteria detect noisy chemical gradients around phytoplankton. Gradient detection depends on bacterial phenotypes and phytoplankton size: large phytoplankton produce spatially extended but shallow gradients, whereas small phytoplankton produce steeper but spatially more confined gradients. To date, it has remained unclear how phytoplankton size and bacterial swimming speed affect bacteria's gradient detection ability and search times for phytoplankton.
Here, we compute an upper bound on the increase in bacterial encounter rate with phytoplankton due to chemotaxis over random motility alone.
We find that chemotaxis can substantially decrease search times for small phytoplankton, but this advantage is highly sensitive to variations in bacterial phenotypes or phytoplankton leakage rates. By contrast, chemotaxis towards large phytoplankton cells reduces the search time more modestly, but this benefit is more robust to variations in search or environmental parameters.
Applying our findings to marine phytoplankton communities, we find that, in productive waters, chemotaxis towards phytoplankton smaller than \SI{2}{\micro\m} provides little to no benefit, but can decrease average search times for large phytoplankton ($\sim\SI{20}{\micro\m}$) from two weeks to two days, an advantage that is robust to variations and favors bacteria with higher swimming speeds. By contrast, in oligotrophic waters, chemotaxis can reduce search times for picophytoplankton ($\sim\SI{1}{\micro\m}$) up to ten-fold, from a week to half a day, but only for bacteria with low swimming speeds and long sensory timescales. This asymmetry may promote the coexistence of diverse search phenotypes in marine bacterial populations.
\end{abstract}

\maketitle

\textbf{Introduction.}
Chemotaxis, the ability to navigate chemical gradients, is often used by marine bacteria to navigate towards phytoplankton cells~\cite{raina2022chemotaxis} and can be important for establishing symbiotic relationships and favoring metabolic exchanges which lie at the heart of the oceans' carbon cycles~\cite{azam1998microbial,raina2019role}.
In the water column, phytoplankton cells generate chemical gradients by leaking dissolved organic compounds that can be strong attractants for bacteria~\cite{thornton2014dissolved,smriga2016chemotaxis}. The region immediately surrounding a phytoplankton cell where organic compounds are present in higher concentration, known as the phycosphere, acts as an ecological interface for the interactions between phytoplankton and bacteria~\cite{seymour2017zooming}.
The composition of marine phytoplankton communities, which span more than two orders of magnitude in cell size from $\sim\SI{0.5}{\micro\m}$ to hundreds of micrometers~\cite{sieburth1978pelagic}, exposes bacteria to gradients on a wide range of lengthscales and amplitudes.

Swimming bacteria perform chemotaxis by sensing temporal changes in the concentration of chemoattractants and biasing their motility towards regions where chemoattractant concentrations are higher~\cite{macnab1972gradientsensing,brown1974temporal,bi2018stimulus}. Because it is based on a molecule-counting process, chemotaxis is inherently subject to noise~\cite{berg1977physics,tenwolde2016fundamental}, especially at the low attractant concentrations typical of marine environments~\cite{lee1975amino}.
The limits that noise imposes on chemosensing are well characterized for concentration fields with large spatial extent~\cite{mora2010limits}, most similar in the ocean to those generated by large phytoplankton, and for strong but short-lived pulses, representative for example of cell lysis events in the sea~\cite{hein2016physical,brumley2019bacteria}.
By contrast, our understanding of bacterial chemotaxis towards small phytoplankton cells, which generate sharp gradients tightly confined in space, is very limited, despite the disproportionate abundance in the ocean of small compared to large phytoplankton cells~\cite{sprules1986plankton,cermeno2008species}. For instance, in a typical phytoplankton community in oligotrophic waters, more than $95\%$ of the population is in the size range between $0.5$ and \SI{3}{\micro\m}, and considerably less than $1\%$ of phytoplankton cells are larger than \SI{10}{\micro\m}.
Historically, less attention has been reserved to chemotaxis towards small phytoplankton after a seminal computational study, though based on parameters determined for the enteric bacterium \textit{Escherichia coli}, established that the gradients generated by phytoplankton cells with radii smaller than $3$--$\SI{4}{\micro\m}$ could not be sensed by a swimming bacterium~\cite{jackson1987simulating}. This view has been recently overturned by NanoSIMS experiments, which revealed that chemotaxis confers an increased nutrient uptake to the marine bacterium \emph{Marinobacter adhaerens} in the presence of the picocyanobacterium \emph{Synechococcus}~\cite{raina2023chemotaxis}, demonstrating that bacterial chemotaxis towards the smallest and most abundant phytoplankton cells in the ocean is possible. Additionally, the swimming speed of bacteria affects both the signal and the noise in their measurement of chemical gradients~\cite{son2016speeddependent} and spans more than one order of magnitude ($10-\SI{100}{\micro\m\per\second}$) in marine bacteria~\cite{grossart2001bacterial,johansen2002variability,kiorboe2002mechanisms}. These results underscore the importance of quantifying the impact of phytoplankton size and bacterial swimming speed on bacterial chemotactic performance, which we address in this study.

\begin{figure*}[t!]
    \centering
    \includegraphics[width=11.4cm]{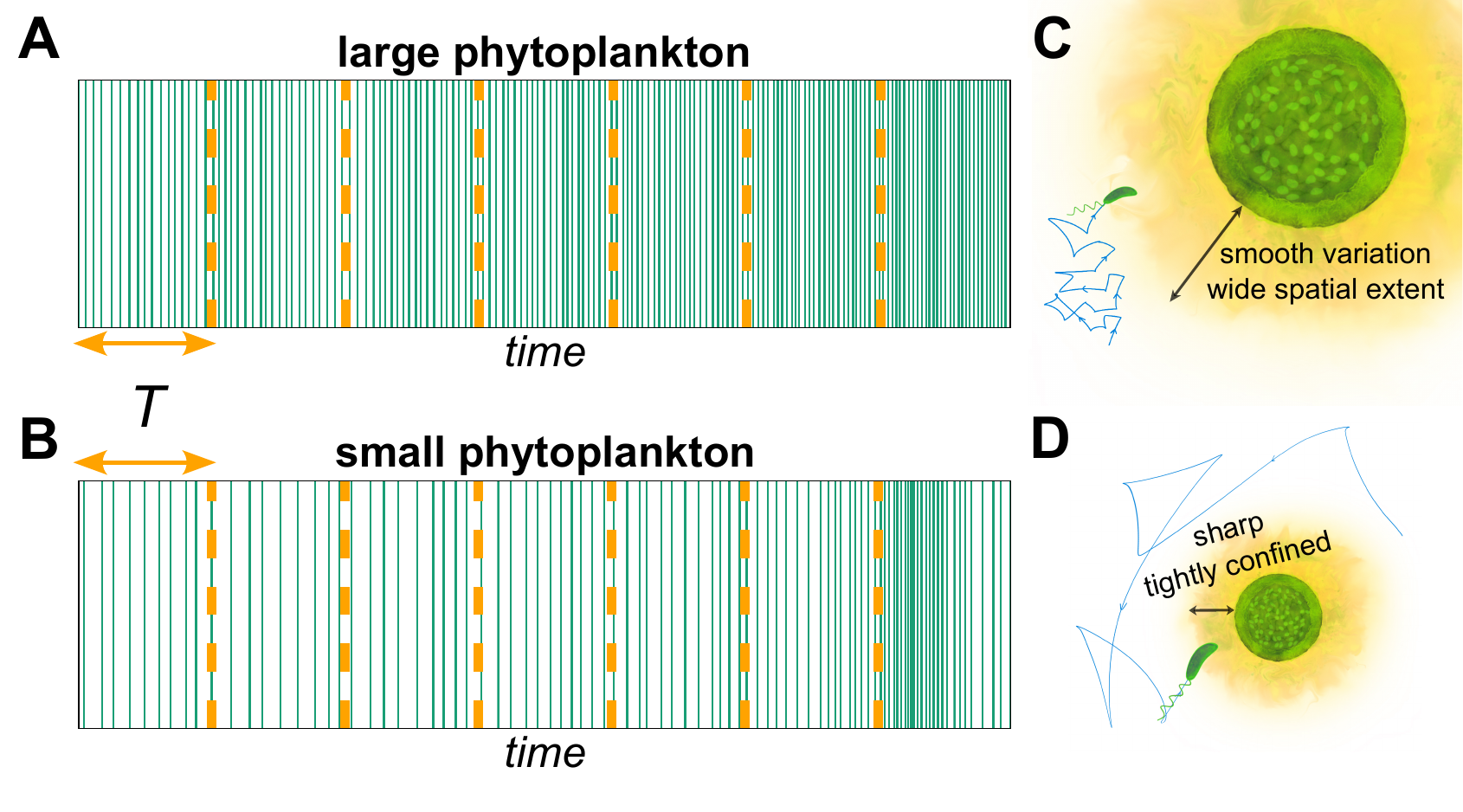}
    \caption{
        \textbf{
        Detecting chemical gradients to increase encounter rates with phytoplankton is a size-dependent challenge for bacteria.
        }
        Bacteria experience concentration gradients in the form of temporal sequences of molecular adsorption events (green bars in panels A, B). Gradients are estimated by integrating the adsorption sequences over intervals of length $T$, the sensory timescale (orange broken lines). 
        The phytoplankton size determines the lengthscale over which the gradients extend (C, D), leading to adsorption sequences with well-distinguished features for small and large sources.
        Large phytoplankton produce profiles with wide spatial extent: the rate of molecular adsorptions increases slowly over time and their detection is limited by molecular fluctuations (A,C).
        Small phytoplankton produce spatially confined profiles: the rate of molecular adsorptions increases sharply over a short time and their detection is limited by the dynamic noise arising from bacterial motion and finite temporal resolution (B,D).
        Timeseries were generated from simulations of Poissonian adsorption events for a bacterium moving at constant speed in the steady-state diffusive concentration field (\autoref{eq:chemoattractant_field}) produced by phytoplankton cells of different sizes.
    }
    \label{fig:setup}
\end{figure*}

\textbf{The role of phytoplankton size in chemotactic searches.}
Phytoplankton cells of different sizes generate chemical gradients of different steepness and spatial extent, posing fundamentally different gradient detection challenges for bacteria.
Bacteria experience chemical concentration fields in the form of temporal sequences of molecular adsorption events, which they integrate over a sensory timescale $T$ to form an estimate of the local concentration gradient~\cite{berg1977physics}. An intuitive understanding of the problem can be obtained by comparing the adsorption sequences experienced by a bacterium moving in the chemoattractant fields generated by a large and a small phytoplankton cell (\autoref{fig:setup}; see also discussion in SI). When moving in the chemoattractant field generated by a large phytoplankton cell, which is typically characterized by a large spatial extent and a shallow gradient, a bacterium will experience a large baseline adsorption rate with only a modest increase over subsequent sensory windows $T$ as it nears the phytoplankton (\autoref{fig:setup}A,C). Detection of such gradients can fail if the gradient is too shallow and gets masked by the fluctuations inherent in the molecular adsorption events~\cite{mora2010limits}.
By contrast, in the case of a small phytoplankton cell the concentration field is typically weaker and tightly localized in space: the signal will thus be indistinguishable from the background until the bacterium is in very close proximity to the phytoplankton, when it will experience a sharp increase in the rate of molecular adsorption events within a short time (\autoref{fig:setup}B,D). Detection of such gradients can fail as a result of the dynamic noise which arises from the combined effect of bacterial motion and the finite sensory timescale $T$: even though the gradients are steep and much less masked by fluctuations, they might not be detected because they occur over timescales smaller than the time required by the bacterium to process the signal. Such resolution limitations are inherent to any measurement system characterized by a finite processing time~\cite{blackman1958measurement}, including the bacterial chemotaxis pathway which is known to act as a low-pass filter~\cite{block1982impulse,tu2008modeling}, but their effect on bacterial chemotaxis towards phytoplankton has remained unexplored. Furthermore, not only does the type of noise change as phytoplankton size decreases, but so does the nature of random encounters between bacteria and phytoplankton: bacteria-phytoplankton encounters are diffusive for large phytoplankton (i.e., the encounter rate scales linearly with phytoplankton size) but ballistic for small phytoplankton (i.e., the encounter rate scales quadratically with phytoplankton size) ~\cite{kiorboe2008mechanistic,slomka2023encounter}.

Here we explore how these limits of gradient sensing, arising from two different types of noise, one associated with inherent fluctuations and the other with the movement and sensory timescale of bacteria, determine chemotactic performance of bacteria towards phytoplankton cells. Combining the limitations on sensing with the ballistic or diffusive nature of random encounters with phytoplankton, we compute an upper bound on the chemotactic index, a dimensionless number that measures the increase in the encounter rates of bacteria with phytoplankton cells due to chemotaxis over random motility alone. Our analysis reveals an asymmetric performance landscape as a function of phytoplankton size. We find that for large phytoplankton, the chemotactic index has a weak dependence on leakage rate and bacterial phenotypes associated with motility and sensing. In stark contrast, for small phytoplankton the chemotactic index is highly sensitive to leakage rate and bacterial phenotypes. 
When considered in the context of encounters within marine phytoplankton communities, our findings reveal that bacteria with low swimming speed and long sensory timescales may obtain large benefits from chemotaxis in the search for small phytoplankton, whereas fast swimmers are unable to exploit chemotaxis in the search for small phytoplankton but perform consistently better in the search for larger phytoplankton.

\begin{figure*}[t!]
    \centering
    \includegraphics[width=17.8cm]{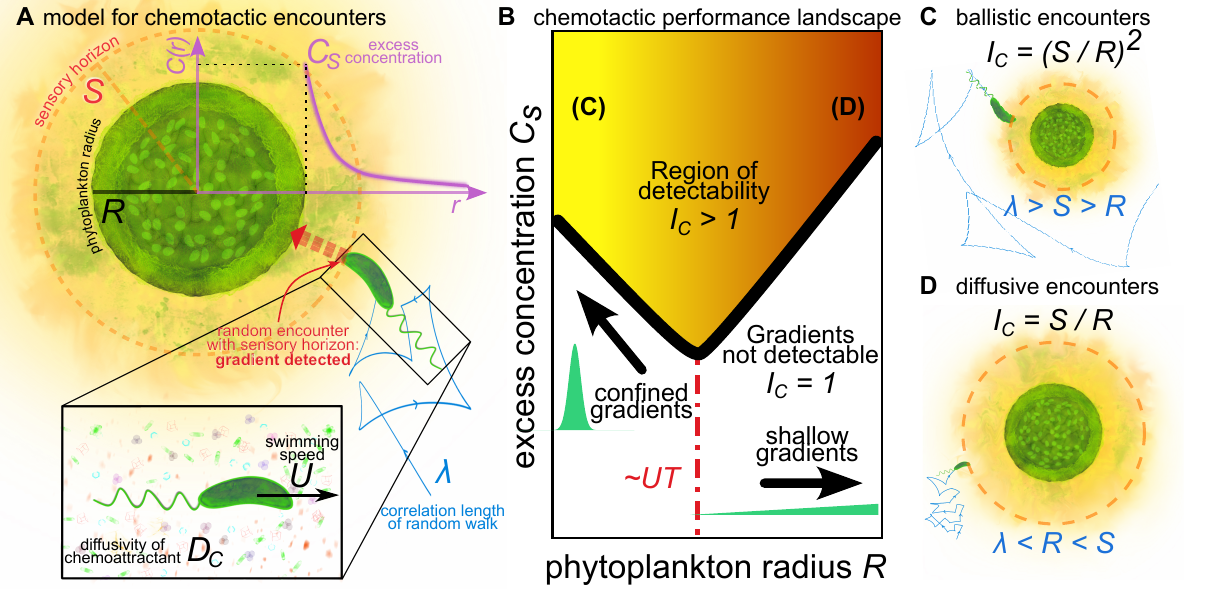}
    \caption{
        \textbf{
        Chemotactic encounters are limited by the gradients' steepness and spatial extent, and ballistic or diffusive encounters with the sensing horizon.
        }
        (A) A phytoplankton cell of radius $R$ produces a chemoattractant field $C(r)=C_0+C_SR/r$ (\autoref{eq:chemoattractant_field}, magenta curve). $C_S$ is the chemoattractant concentration at the phytoplankton surface in excess of the bulk background concentration $C_0$.
        Far from the cell, bacteria cannot sense the chemoattractant and thus swim in unbiased random walks, with correlation length $\lambda$.
        The ``sensory horizon'' $S$ is the distance from the phytoplankton cell at which a perfect chemotaxer can detect the gradient and encounters its target with 100\% probability~(\autoref{eq:snr}).
        (Inset) Chemosensing is a molecule-counting process driven by the adsorption of individual chemoattractant molecules, which reach the bacterium via diffusion (we consider a single chemical species with diffusivity $D_C$) while the bacterium swims at speed $U$.
        (B) Depiction of the performance landscape for chemotactic searches.
        The phytoplankton radius $R$ and the excess chemoattractant concentration $C_S$, which we treat as independent parameters both here and in Figures~\ref{fig:asymmetric-performance} and \ref{fig:strategies}, determine whether gradients can be successfully detected or not.
        For a given phytoplankton radius $R$, there is a minimum value of $C_S$ for gradient detection to be possible; this set of values defines a convex boundary of detection (thick black line) which separates regions in the landscape where gradient detection is possible (the region above) or not (the region below).
        Below the boundary of detection, two conditions limit an organism's ability to perform chemotaxis: gradients are either too spatially confined (left, \autoref{eq:left_boundary}) or too shallow (right, \autoref{eq:right_boundary}). The limiting factor for detection is determined by the relationship between phytoplankton radius $R$ and the distance traveled by a bacterium during one sensory interval, $UT$.
        In the region of detectability ($I_C>1$) $I_C$ is determined by the relationship between the swimming correlation length $\lambda$, the phytoplankton radius $R$ and the sensory radius $S$.
        The size dependence of random encounters identifies two subregions within the region of detectability (here qualitatively represented by the yellow-red shading), in which the chemotactic index displays two distinct behaviors.
        Small phytoplankton (C) lead to ballistic encounters for which the chemotactic index scales quadratically with the sensory radius $S$, whereas for large phytoplankton (D) the chemotactic index scales only linearly with $S$ due to the diffusive nature of encounters.
    }
    \label{fig:sensing}
\end{figure*}

\textbf{Theoretical model for chemotactic encounters.}
To understand how phytoplankton size affects the ability of bacteria to detect gradients, we consider an ideal bacterium whose chemotactic performance is limited only by the ability to detect a chemoattractant gradient at a distance from a phytoplankton cell. Following previous work~\cite{jackson1987simulating,seymour2017zooming}, we represent a phytoplankton cell as a sphere of radius $R$ (\autoref{fig:sensing}A) that continuously exudes chemoattractant homogeneously through its surface, via either active or passive mechanisms~\cite{seymour2017zooming}. Regardless of the exudation mechanism, the chemoattractant then diffuses away from the cell. At steady state, which is reached within timescales of seconds to minutes (SI), the concentration field around the cell is
\begin{equation}\label{eq:chemoattractant_field}
    C(r) = C_0 + C_S \dfrac{R}{r},
\end{equation}
where $C_0$ is the background concentration of the chemoattractant far from the phytoplankton, $C_S$ is the excess concentration of the chemoattractant at the phytoplankton cell surface (i.e., the concentration above $C_0$), and $r$ is the radial distance from the center of the phytoplankton cell. In the SI we show that all results obtained below are also valid if bacterial consumption of chemoattractant is taken into account, leading to an exponentially screened concentration field.
In the absence of chemoattractants, a motile bacterium performs a random walk~\cite{berg1993random} with swimming speed $U$ and correlation length $\lambda$ (\autoref{fig:sensing}A). When chemoattractant gradients are present, the bacterium can bias its motion up the gradient, increasing its chances of encountering the phytoplankton.
An exact evaluation of the encounter-enhancing effect of chemotaxis requires explicit consideration of the details of the bacterial motility and chemosensory system, which may vary significantly across species and physiological states. With the aim of seeking more general conclusions, we focus instead on computing an upper bound to the increase in encounter rates afforded by chemotaxis that will be valid for a wide variety of organisms, by considering the bacterium to be a ``perfect chemotaxer''.
We assume that, around a phytoplankton cell, there is a sensory horizon $S$, which corresponds to the maximal distance at which a bacterium can reliably detect the gradient generated by the phytoplankton (\autoref{fig:sensing}A). Outside the sensory horizon no chemotaxis is possible and the bacterium moves purely by random motility, but as soon as the sensory horizon is reached, the perfect chemotaxer is able to navigate flawlessly towards the phytoplankton cell, always yielding an encounter with the phytoplankton. This concept of sensory horizon can be considered equivalent to those of capture radius, sensing range or reaction distance often introduced in the study of predation and encounters in higher organisms~\cite{utne1997effect,hutchinson2007use,visser2007motility,andersen2016characteristic}. The encounter rate of the perfect chemotaxer is thus limited only by random encounters with the sensory horizon $S$: this simplification of the problem allows one to use existing solutions for random encounters between spherical objects~\cite{slomka2023encounter}.
The problem of chemotactic encounters with the phytoplankton cell is therefore reduced to the problem of identifying the sensory horizon $S$, as a function of the phytoplantkon radius $R$.
Due to its idealized response, which always ensures an encounter with the phytoplankton upon gradient detection, the performance of the perfect chemotaxer is an upper limit on the performance of any real chemotactic bacterium.

\textbf{The signal-to-noise ratio determines the effectiveness of chemotaxis.}
The spatial extent of the sensory horizon is defined as the distance from the phytoplankton at which bacteria can detect the gradient of chemoattractants exuded by the phytoplankton. We determine this distance in terms of the signal-to-noise ratio (SNR) associated with the bacterial measurements of the gradient, extending the approach Hein et al.~\cite{hein2016physical} used to study chemotaxis towards ephemeral Gaussian pulses, by including the noise associated with the spatial confinement of the signal. In a steady concentration profile, the signal is the rate of change of concentration experienced by the bacterium over time as it swims, $|U\nabla C|$~\cite{berg1977physics,hein2016physical}. The noise arises through the fluctuations in the adsorption of attractant molecules, which reach the bacterial cell surface with diffusivity $D_C$ (\autoref{fig:sensing}A inset)~\cite{berg1977physics}.
In constant gradients (i.e., concentration fields varying linearly with distance from their source) the inherent noise in the gradient measurement arising from fluctuations in molecular adsorption events was derived by Mora \& Wingreen~\cite{mora2010limits} as $\sigma_0 = \sqrt{3C/(\pi a D_C T^3)}$.
The gradient associated with the concentration field around a phytoplankton cell (\autoref{eq:chemoattractant_field}) is instead not constant, having a magnitude $|\nabla C(r)| = C_S R/r^2$ that increases as bacteria approach the phytoplankton and attains its maximum at the phytoplankton surface.
As a first step to account for this increase, we approximate the local gradient experienced by the bacterium within a single sensory window $T$ as a linearly increasing gradient, which yields the following revised estimate of the noise (SI),
\begin{equation}
\label{eq:noise_2ndorder}
  \sigma_0^* = \sigma_0\sqrt{1 + \dfrac{3}{20}\Big(\dfrac{UT}{R+\Delta r}\Big)^2},
\end{equation}
where $\Delta r$ is the distance from the surface of the phytoplankton cell.
\eqref{eq:noise_2ndorder} highlights the importance of bacterial movement and phytoplankton size in the sensing process: for large phytoplankters ($R\gg UT$), at any distance $\Delta r$ the correction to $\sigma_0$ is negligible; but when the phytoplankton is small compared to the distance traveled by the bacterium during the sensory interval $T$ and the bacterium is close to the phytoplankton ($UT > R+\Delta r$), the sensing noise increases considerably.
This result shows that motion in non-linear concentration fields has the effect of a low-pass filter, which prevents bacteria from accurately measuring high-frequency variations in the concentration field, i.e., gradients tightly confined in space.
Interestingly, the equation for the sensing noise highlights a tradeoff in the bacterial swimming speed: while a larger speed $U$ increases the signal (the local gradient $|U\nabla C|$), it also increases the sensing noise in the detection of gradients generated by small phytoplankton (\eqref{eq:noise_2ndorder}).
Higher swimming speeds may, therefore, not always be beneficial for chemotaxis to phytoplankton, in particular when phytoplankton cells are small.

While it clarifies the role of movement and size, the approximation of the gradient as locally linear, that we used to derive \eqref{eq:noise_2ndorder}, does not yet capture the limit of the smallest phytoplankton cells, for which the sharp increase in the gradient amplitude may be so tightly confined that it occurs within a single sensory interval $T$ (Fig.~\ref{fig:setup}B,D).
This is ecologically a very important scenario, yet we found this case to be analytically intractable. We therefore introduce, phenomenologically, a low-pass filter $f(x)=1-\exp(-x^{3/2})$ in the expression of the $\SNR$, which degrades the high-frequency components of the signal ($f(x\gg1)\sim0$) without affecting the detection of lower frequency components ($f(x\ll1)\sim1$).
We then define the sensory horizon $S$ as the farthest distance from the phytoplankton cell at which the $\SNR$ is above a threshold $q$. $S$ is, therefore, the solution to the equation
\begin{equation}\label{eq:snr}
  \SNR = \dfrac{|U\nabla C(S)|f(UT/S)}{\Pi\sigma_0(S)} = q,
\end{equation}
which we solve numerically~(SI). 
In \autoref{eq:snr} we introduced a constant chemotactic precision factor $\Pi$ to explicitly represent noise amplification in the signal processing internal to the chemotaxis pathway~\cite{brumley2019bacteria}. This choice is motivated by previous observations that in the marine bacterium \textit{Vibrio anguillarum} the response to chemoattractant pulses was accurately described by multiplying the theoretical noise, $\sigma_0$, by a chemotactic precision factor $\Pi\approx6$~\cite{brumley2019bacteria}. In what follows, we therefore set $\Pi=6$.
We also set $q=1$ throughout, which equates to defining $S$ as the distance from the phytoplankton where bacteria can detect a positive gradient with a probability of $\approx 84\%$~(SI).
While a precise numerical solution of \autoref{eq:snr} will be obtained later, an approximate calculation readily determines the boundary of detection, i.e., a curve $C_{S,\text{min}}(R)$ in the $(R,C_S)$ performance landscape that determines whether a phytoplankton of radius $R$ is detectable ($C_S>C_{S,\text{min}}(R)$) or not ($C_S<C_{S,\text{min}}(R)$). By solving \autoref{eq:snr} in the two limit cases $S=UT$ and $S=R$, corresponding respectively to the detection being limited by gradient confinement and gradient steepness, and with the additional approximation $C_0=0$, we obtain two curves (SI)
\begin{subequations}\label{eq:boundaries}\begin{align}
    &C_{S,\text{min}}^\text{left} \approx 34.4\dfrac{U}{aD_CR}, &\quad (S=UT) \label{eq:left_boundary} \\
    &C_{S,\text{min}}^\text{right} \approx 34.4\dfrac{R^2}{aD_CU^2T^3}, &\quad (S=R) \label{eq:right_boundary}
\end{align}\end{subequations}
which define a convex region in the space of phytoplankton phenotypes $R$ and $C_S$ (\autoref{fig:sensing}B), that we call the region of detectability, where bacteria can detect the gradients produced by phytoplankton cells.

When gradient detection is possible, the performance of a bacterium's search for a phytoplankton is limited by its random encounters with the sensory horizon, since no chemotaxis is possible outside the sensory horizon.
For perfect chemotaxis, reaching the sensory horizon $S$ ensures that the bacterium will eventually reach the phytoplankton cell: the problem of chemotactic encounters with the phytoplankton cell thus simplifies to the problem of random encounters with the sensory horizon.
We can then quantify the maximum possible performance of a chemotactic search through the chemotactic index, $I_C$, which we define as the ratio between the rate of random encounters with the sensory horizon $S$ and the rate of random encounters with the phytoplankton cell of radius $R$, as
\begin{equation}\label{eq:IC}
  I_C = \Gamma(S)/\Gamma(R).
\end{equation}
$\Gamma(x)$ is the encounter kernel that quantifies the random encounters of the bacterium with a target of radius $x$ and represents the amount of volume swept per unit time by the motion of the bacterium relative to the phytoplankton~\cite{hutchinson2007use,kiorboe2008mechanistic,slomka2023encounter}. For simplicity, we only consider nondestructive encounters~\cite{gurarie2013general}, in which the phytoplankton and chemical gradients remain unaffected by the encounter event. Moreover, we ignore stages of interaction successive to the encounter with the sensory horizon, after which the bacterium might, for example, reside in the proximity of the phytoplankton for prolonged times~\cite{raina2023chemotaxis} and experience multiple successive encounters with the same phytoplankton. Chemosensing is successful and increases encounters when the bacterium can detect a gradient at a distance from the phytoplankton cell, that is when a sensory horizon $S>R$ exists; in this case, the chemotactic index will be $I_C>1$.
When a sensory horizon $S>R$ does not exist, bacteria cannot detect gradients at a distance from the phytoplankton cell, so chemotaxis cannot enhance encounter rates and therefore $I_C=1$ (we ignore possible scenarios where chemotaxis is detrimental to encounters, leading to $I_C<1$).
Within the region of detectability ($I_C>1$, \autoref{fig:sensing}B), $I_C$ is determined by the relationship between the correlation length of the bacterial random walk, $\lambda$, the phytoplankton radius, $R$, and the sensory horizon, $S$ (SI, Equation S42). In particular, the encounters with small phytoplankton cells ($\lambda>S>R$) have a ballistic nature ($\Gamma\sim R^2$)~\cite{maxwell1860illustrations} and result in a quadratic scaling of the chemotactic index with the sensory horizon, $I_C = S^2/R^2$ (\autoref{fig:sensing}C). For large phytoplankton cells ($\lambda < R < S$) encounters have a diffusive nature ($\Gamma\sim R$)~\cite{smoluchowski1916drei,chandrasekhar1943stochastic} and the chemotactic index scales only linearly with the sensory horizon, $I_C = S/R$ (\autoref{fig:sensing}D).

\begin{figure*}[t!]
    \centering
    \includegraphics[width=17.8cm]{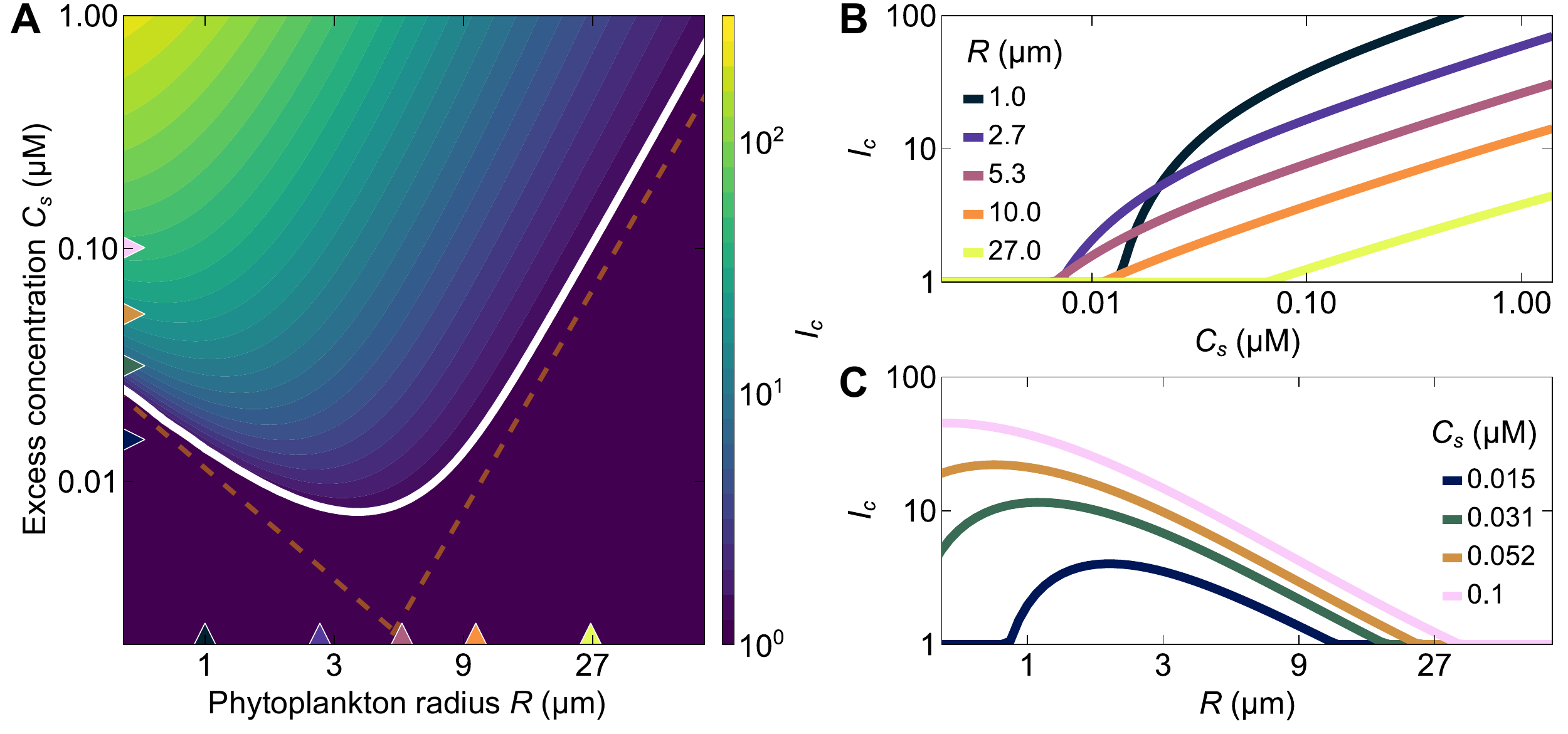}
    \caption{
        \textbf{
        The stakes are high for encounters with small phytoplankton -- the chemotactic index is largest for small phytoplankton, but it drops sharply when gradient detection fails.
        }
        (A) Performance ($I_C$) landscape for a bacterium using chemotaxis to drive encounters with spherical targets of different radii ($R$) and chemoattractant concentrations ($C_S$) calculated using \autoref{eq:snr} and \autoref{eq:IC} (see also Equation~(S42)).
        The thick white line defines the boundary of detection, separating the region where gradients are detectable and chemotaxis is beneficial to encounters ($I_C > 1$) from the region where gradients are too shallow or too spatially confined to be detected ($I_C = 1$). Broken orange lines are the approximate predictions for the minimum value of $C_S$ at which detection is possible (\autoref{eq:boundaries}).
        (B) Vertical transects from panel A for fixed values of the target radius $R$ (corresponding to the upward-pointing triangles).
        In chemotaxis towards small targets, a slight variation in the chemoattractant concentration can make the difference between a highly successful search ($I_C\sim10$) or a failure ($I_C=1$), whereas for larger targets the dependency of $I_C$ on the chemoattractant concentration is more gradual.
        (C) Horizontal transects from panel A for fixed values of the chemoattractant concentration $C_S$ (corresponding to the right-pointing triangles).
        For weak sources, an increase in size $R$ produces initially large enhancements in chemotactic performance, with diminishing returns upon further enlargement.
        As the source gets stronger, the increase in performance for chemotaxis towards small targets becomes disproportionately larger; larger sources also become detectable although offering modest performance improvements over random searches.
        A script to generate an interactive dashboard for the rapid evaluation of the $I_C$ landscape is available from \href{https://github.com/mastrof/GradientSensing/blob/main/dashboards/IC_landscape.jl}{GitHub} (see Movie S1 for a demo).
    }
    \label{fig:asymmetric-performance}
\end{figure*}

We note that our definition of the chemotactic index is analogous to that used in the In-Situ Chemotaxis Assay (ISCA). The ISCA is a microfluidic assay for deployment in aqueous environments. It measures the strength of bacterial chemotaxis to a given compound in situ by comparing the number of bacteria that accumulate, after a given deployment time of typically \SI{1}{hour}, in a well filled with that compound, to the number of bacteria found in a negative control well devoid of chemoattractants~\cite{lambert2017microfluidicsbased,raina2022chemotaxis,clerc2023strong}.
Our results may, therefore, be interpreted as providing an upper bound on the chemotactic index obtained in ISCA experiments with inlets of different sizes containing chemoattractants in different concentrations.
We next solve \eqref{eq:snr} numerically for $S$ and use the solution to compute $I_C$ according to \eqref{eq:IC} and thus quantify the chemotactic performance as a function of bacterial and phytoplankton phenotypes.

\textbf{Asymmetry of chemotactic performance.}
We find that chemotaxis towards small phytoplankton cells can be risky but highly rewarding.
We illustrate this by applying our model to a typical motile marine bacterium with radius $a=\SI{0.5}{\micro\m}$, swimming speed $U=\SI{50}{\micro\m/\s}$ and sensory timescale $T=\SI{100}{\milli\s}$, that is chemotactic towards a low-molecular-weight compound with diffusivity $D_C=\SI{500}{\micro\m^2/\s}$ (\autoref{fig:asymmetric-performance}). We assume the background concentration of the compound ($C_0$ in \autoref{eq:chemoattractant_field}) to be \SI{1}{\nano M}, typical of oligotrophic ocean waters~\cite{lee1975amino} (calculations with different background concentrations are shown in Figure S6). Using our model, we quantified the chemotactic performance landscape by computing the chemotactic index as a function of the phytoplankton radius $R$ and the excess concentration $C_S$ (which is proportional to the phytoplankton leakage rate). We stress that, in the analysis below, $C_S$ and $R$ are free parameters; later, we will include the physiological constraints the cell size imposes on the excess concentration. The performance landscape exhibits a convex, V-like, shape, with its apex around $R \approx UT$, as predicted by the approximate \autoref{eq:boundaries}, and shows a marked asymmetry between small and large phytoplankton (\autoref{fig:asymmetric-performance}A). Specifically, the chemotactic index has a much sharper dependency on $C_S$ when the phytoplankton radius $R$ is small ($R<\SI{2}{\micro\m}$ in this example), especially close to the boundary of detection ($I_C=1$) (\autoref{fig:asymmetric-performance}B). This indicates that even modest variations in the excess concentration $C_S$ leaked by a small phytoplankton can have large impacts on the performance of bacterial chemotaxis. In the example, while bacteria will not be able to sense ($I_C=1$) an attractant gradient from a \SI{1}{\micro\m} phytoplankton creating an excess concentration of $C_S=\SI{10}{\nano M}$, a moderately higher value $C_S=\SI{30}{\nano M}$ yields a high chemotactic index $I_C=10$. As $C_S$ is further increased, the dependency of $I_C$ on $C_S$ becomes weaker.
For larger phytoplankton cells, the increase in performance is more gradual: above the boundary of detection ($I_C=1$), $I_C$ grows slowly with increasing $C_S$.
Looking across phytoplankton radii $R$, for small values of the excess concentration $C_S$ an increase in $R$ leads first to a sharp increase in chemotactic index and then a gradual decrease towards $I_C=1$ when $R$ exceeds an optimal value (\autoref{fig:asymmetric-performance}C).

The behavior of the chemotactic index close to the boundary of detection ($I_C=1$) reveals the origin of the asymmetry in the chemotactic performance landscape.
For chemotaxis towards large phytoplankton (right boundary), gradient detection is limited by gradient steepness. If the excess concentration of chemoattractant around the phytoplankton is just sufficient to allow gradient detection, the sensory horizon will only be marginally bigger than the size of the phytoplankton, $S = R(1+\varepsilon)$ (where $\varepsilon\ll 1$), because the gradients in the concentration profile in \autoref{eq:chemoattractant_field} are steepest near the phytoplankton. Since encounters with large targets are diffusive, the encounter rate is proportional to target size, yielding $I_C \sim S/R \sim 1 + \varepsilon$: the chemotactic index increases smoothly as the boundary of detection is crossed.
By contrast, for small phytoplankton (left boundary), gradient detection is limited by the gradient's spatial extent. The minimal sensing horizon that can be detected is therefore on the order of the distance traveled by the bacterium in a single sensory interval, $S\sim UT$. Since encounters with small cells are ballistic, the encounter rate is proportional to the square of the target size, and thus $I_C \sim (S/R)^2 \sim (UT/R)^2$: the chemotactic index exhibits a large jump from 1 to $(UT/R)^2$ as the boundary is crossed.
For example, in \autoref{fig:asymmetric-performance} we have $UT=\SI{5}{\micro\meter}$, which implies
a jump of the order $(UT/R)^2 \sim 25$ for a small phytoplankton cell with
$R=\SI{1}{\micro\meter}$, which captures the order of magnitude
of the jump observed in \autoref{fig:asymmetric-performance}C (green line). 
We next show that this asymmetry of the performance landscape is robust to changes in the key parameters.

\begin{figure*}[t!]
    \centering
    \includegraphics[width=17.8cm]{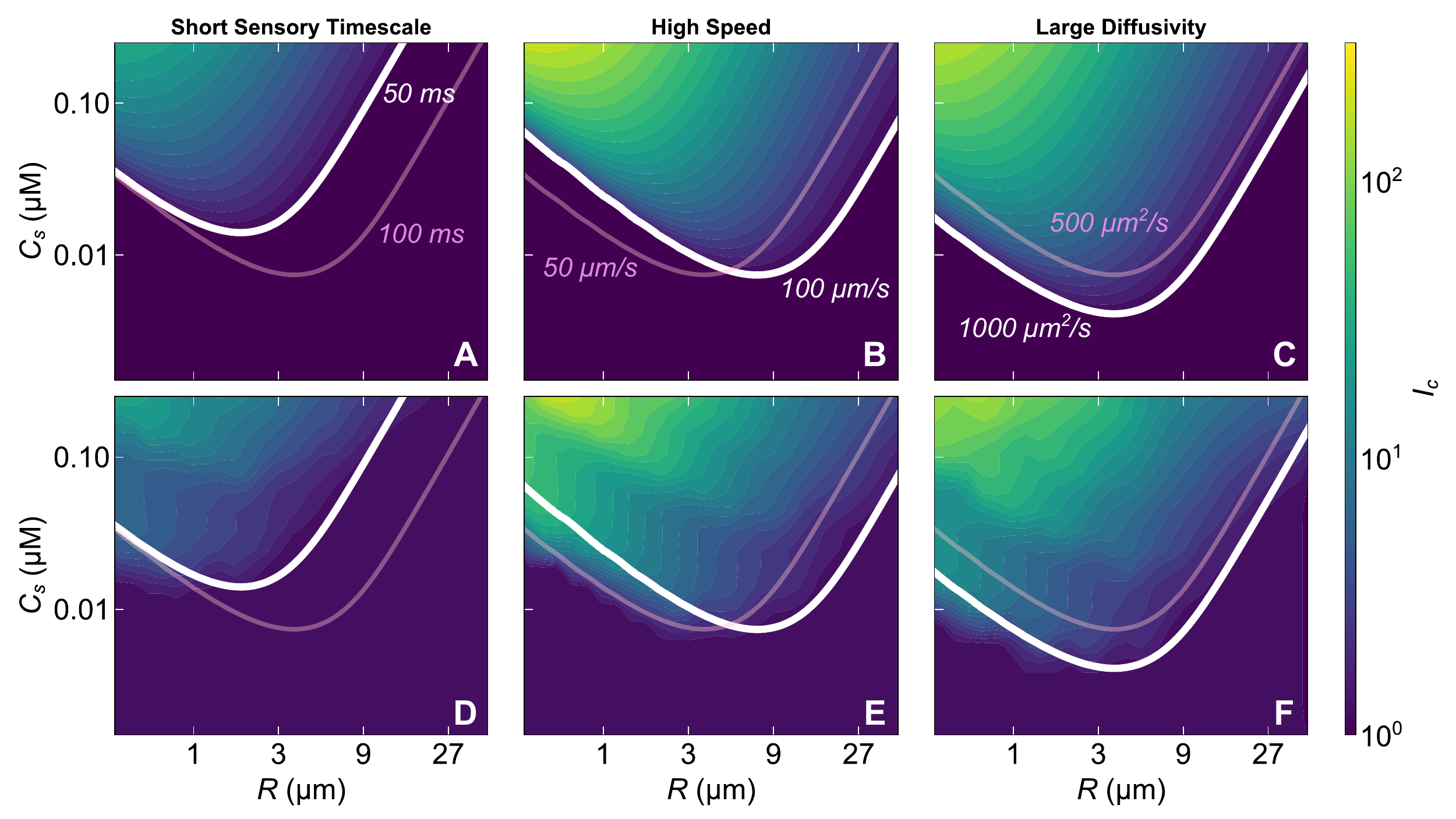}
    \caption{
        \textbf{
        Bacterial phenotypes and chemoattractant diffusivity control the chemotactic index but do not alter the high-stakes nature of encounters with small phytoplankton.
        }
        Performance landscapes of different chemotactic strategies obtained from theoretical (A--C) and computational (D--F) models.
        In all the panels, the pink curve is the detection boundary for a reference system with swimming speed $U=\SI{50}{\micro\m/\s}$, sensory timescale $T=\SI{100}{\milli\s}$ and chemoattractant diffusivity $D_C=\SI{500}{\micro\m/\s^2}$ (same as in \autoref{fig:asymmetric-performance}A).
        (A) Reduction in sensory timescale $T$ from \SI{100}{\milli\s} to \SI{50}{\milli\s}.
        (B) Increase in swimming speed from \SI{50}{\micro\m/\s} to \SI{100}{\micro\m/\s}.
        (C) Increase in chemoattractant diffusivity from \SI{500}{\micro\m^2/\s} to \SI{1000}{\micro\m^2/\s}.
        (D--F) Theoretically predicted features of bacterial chemotactic performance are reproduced by a minimal model of an ideal sensor based on the Kolmogorov-Smirnov test.
        In panels D to F, the white line is the detection boundary from the theoretical prediction of panels A to C, respectively; the heatmap represents values of the chemotactic index obtained from numerical simulations of the ideal sensor (linearly interpolated).
        Despite small quantitative differences in the estimated $I_C$ values, arising from the distinct signal processing mechanism and the finite spatial resolution of numerical simulations,
        the features of the performance landscape (shape and asymmetry) are clearly conserved.
    }
    \label{fig:strategies}
\end{figure*}

\textbf{Dependence of chemotactic performance on physical parameters.}
Variations in the swimming speed $U$, the sensory timescale $T$ and the chemoattractant diffusivity $D_C$, determine the performance of chemotaxis for a given phytoplankton radius and excess concentration, but do not affect the fundamental structure of the performance landscape (\autoref{fig:strategies}).
This can be seen by comparing the landscapes obtained through the variations in individual parameters against the reference landscape computed in \autoref{fig:asymmetric-performance}, where we used $U=\SI{50}{\micro\m/\s}$, $T=\SI{100}{\milli\s}$ and $D_C=\SI{500}{\micro\m^2/\s}$.
A reduction in the sensory timescale $T$ from $100$ to \SI{50}{\milli\s} (two values on the low end of the estimates for bacterial sensory timescales~\cite{berg1977physics,segall1982signal,jackson1987simulating,sourjik2002binding,sagawa2014singlecell}) decreases the ability of bacteria to detect gradients from large phytoplankton cells and reduces the overall performance of chemotaxis (\autoref{fig:strategies}A; Figure S7A and S8D). A smaller value of $T$ is associated with a larger sensing noise ($\propto T^{-3/2}$) but with a smaller dynamic noise because the measurement frequency is increased. The position of the left boundary of detection is unaffected by the variation in $T$ (\autoref{eq:left_boundary}) but the right boundary is shifted towards larger values of $C_S$ as $T$ increases (\autoref{eq:right_boundary}).
An increase in the swimming speed $U$ from $50$ to \SI{100}{\micro\m/\s} (respectively a moderate and a large value for the average swimming speed of marine bacteria~\cite{mitchell1995long,johansen2002variability}) improves the performance of chemotaxis within the region of detectability (since the raw signal $|U\nabla C|$ increases), but the region of detectability itself is shifted towards larger values of the phytoplankton radius: as $U$ increases, higher $C_S$ is required to detect small phytoplankton (\autoref{eq:left_boundary}) but lower $C_S$ is required to detect large phytoplankton (\autoref{eq:right_boundary}). This result highlights that an increased swimming speed results in a tradeoff between degraded spatial resolution, due to the lower frequency of measurements, and improved sensitivity (\autoref{fig:strategies}B; Figure S7B and S8E). A surprising consequence is that lower swimming speeds may improve the performance of chemotactic searches towards small phytoplankton cells.
An increase in the diffusivity of the chemoattractant $D_C$ from $500$ to \SI{1000}{\micro\m^2/\s} 
(values which are representative respectively of sugars~\cite{schramke1999prediction}, and DMSP~\cite{spiese2018determination} or amino acids~\cite{ma2005studies})
enhances the signal-to-noise ratio (sensing noise $\propto D_C^{-1/2}$) without affecting spatial resolution, and thus shifts the region of detectability towards lower $C_S$ values (\autoref{fig:strategies}C; Figure S7C and S8F), as also described by the $1/D_C$ dependency of the two boundaries (\autoref{eq:boundaries}).
In summary, we have shown that bacterial chemotactic searches for phytoplankton cells are characterized by a fundamental asymmetry in the performance, which does not depend on the specific values of the parameters involved in the search: the performance of chemotaxis towards large phytoplankton varies smoothly with variations in leakage rate and other parameters, while the performance of chemotaxis towards small phytoplankton is highly sensitive to variations in the search parameters and may exhibit sudden jumps between high-gain and no gain at all ($I_C=1$). This suggests an interpretation of chemotaxis to small phytoplankton cells as a high-stakes adaptation.

\textbf{An ideal sensor model.}
To demonstrate that our conclusions are not the result of specific model choices, we further develop a more general framework based on a minimal numerical model of an ideal sensor (\autoref{fig:strategies}D--F).
This ideal sensor is defined as a sphere moving at constant speed $U$ towards a phytoplankton cell surrounded by a chemoattractant field $C(r)$.
As it moves, the ideal sensor registers all the adsorption events of chemoattractant molecules, occurring as Poisson events with instantaneous rate $4\pi D_C a C(r)$, where $r$ is the instantaneous distance of the sensor from the center of the phytoplankton. After a time interval $T$, all registered events are processed and a new acquisition starts.
For signal processing, the sensor performs a one-sided Kolmogorov-Smirnov test~\cite{massey1951kolmogorovsmirnov} comparing the distribution of the waiting times between successive adsorption events recorded in the first and the second half of the acquisition interval $T$.
If the cumulative distribution function is larger in the second half of the interval than in the first half (with p-value $p<0.05$), then the sensor has detected a positive gradient. Averaging the successful gradient detections over an ensemble ($N=1000$) of such ideal sensors provides an estimate of the sensory distance $S$, defined as the largest distance $r$ where at least a fraction $f$ of the sensors has detected a gradient. The estimates of $S$ can then be used to evaluate the chemotactic index $I_C$ as we have done for the bacterial model (\autoref{eq:IC}).
Remarkably, we find that with a high consensus threshold ($f=0.99$), the ideal sensor model provides close agreement with our theoretical calculations, both in terms of the V-shape of the detectability region and of the predicted $I_C$ values (\autoref{fig:strategies}D--F). One discrepancy is the exact location of the left boundary of the detectability region, which reflects the sensitivity of the chemotactic index to the details of the signal processing, i.e., to the way high-frequency signals are degraded. In \autoref{eq:snr}, we represented the signal degradation through the phenomenological low-pass filter $f(x)$; the Kolmogorov-Smirnov sensor still shows low-pass characteristics for small phytoplankton but the filtering function is different (for a detailed discussion of this minimal model and the simulation procedure, see SI and Figure S9).
This close agreement highlights the generality of our findings, since they are recapitulated by an ideal sensor lacking many of the specific search characteristics of a chemotactic bacterium.
We next explore the ecological consequences of the asymmetry in the performance of bacterial chemotaxis towards phytoplankton, by taking into account physiologically realistic values of the excess concentration and using the resulting chemotactic indices to estimate typical search times in  phytoplankton communities.

\begin{figure*}[t!]
    \centering
    \includegraphics[width=17.8cm]{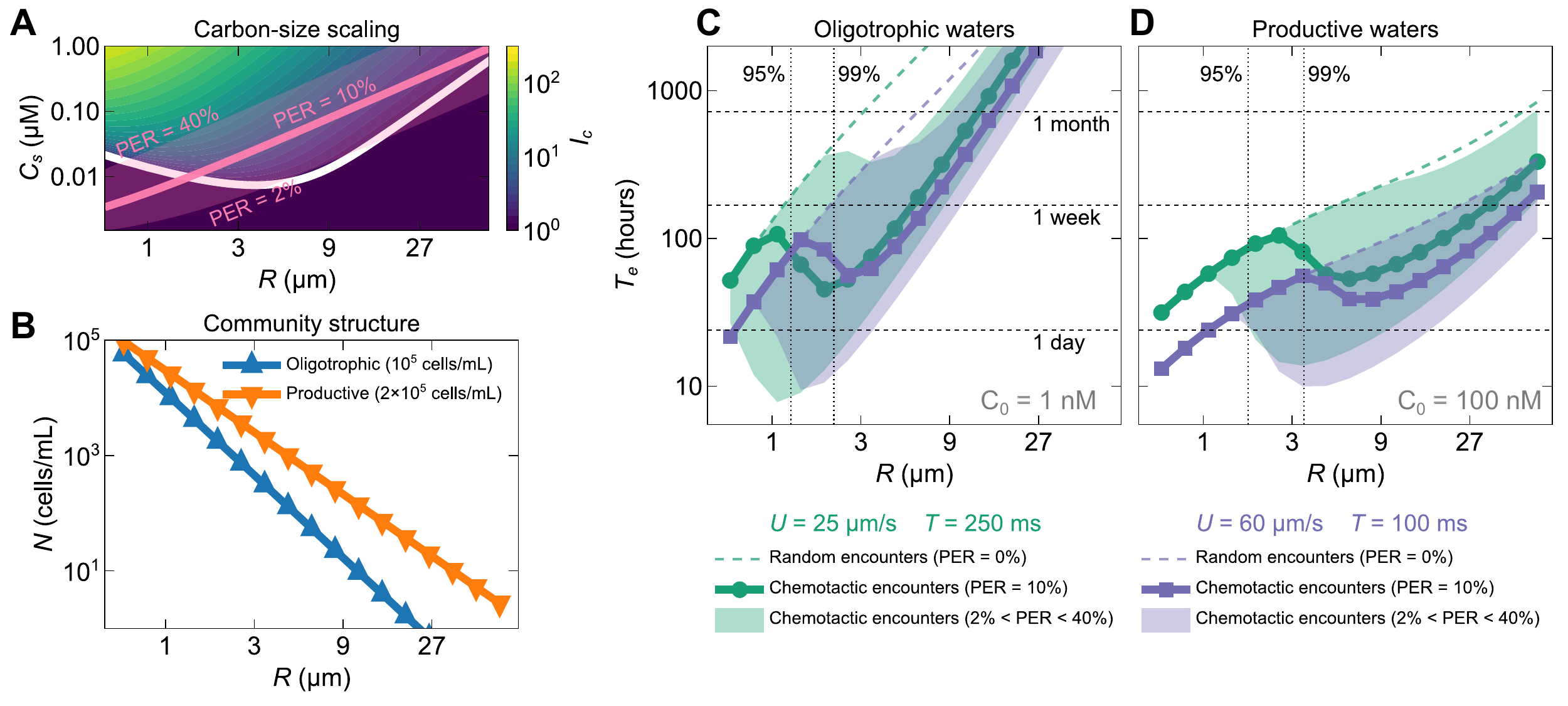}
    \caption{
        \textbf{
        The asymmetric performance of bacterial chemotaxis towards phytoplankton cells may promote a diversity of search phenotypes.
        }
        (A) Characteristic values of size and released chemoattractant for phytoplankton cells are constrained by carbon-size scaling laws (pink areas). When overlaid on the performance landscape of a bacterium (same as in \autoref{fig:asymmetric-performance}A), the physiological range of exudation rates lies across the boundary of detection.
        The thick pink line corresponds to phytoplankton cells with a percent extracellular release (PER) of 10\%, the shaded pink band represents variations in the percent extracellular release between 2\% (lower limit) and 50\% (upper limit).
        (B) Marine phytoplankton communities are dominated by small cells. The size structure follows a power-law distribution, where the abundance $N$ decreases with increasing size $R$ as $N(R) \propto R^{-3\alpha}$.
        The allometric exponent is larger in oligotrophic waters ($\alpha=1.0$) than in more productive waters ($\alpha=0.75$). For the oligotrophic community we assume a total abundance of \SI{1e5}{cells/\milli\liter}, and \SI{2e5}{cells/\milli\liter} for the productive community.
        (C-D) Comparison of average search times between random motility and chemotaxis.
        Carbon-size scaling, phytoplankton community structure and chemotactic index jointly define the average search time ($T_e$) (\autoref{eq:search_time}) an individual bacterium requires to encounter a phytoplankton cell using chemotaxis.
        The thin broken lines represent the search times in the absence of chemotaxis ($I_C=1$).
        The thick lines with markers indicate search times for phytoplankton cells with percent extracellular release of 10\% (corresponding to the thick pink line in panel A), and the shaded bands represent variations in percent extracellular release between 2\% and 40\% (matching the pink shaded band in panel A).
        The two curves correspond to distinct bacterial phenotypes: in green with circle markers, a bacterium with low speed and long sensory timescale; in violet with square markers, a bacterium with high swimming speed and short sensory timescale.
        The vertical dotted lines mark the radii corresponding to the 95\textsuperscript{th} and 99\textsuperscript{th} percentiles of the phytoplankton community abundance.
        A script to generate an interactive dashboard for the rapid evaluation of search times is available from \href{https://github.com/mastrof/GradientSensing/blob/main/dashboards/search_times.jl}{GitHub} (see Movie S2 for a demo).
    }
    \label{fig:ecology}
\end{figure*}

\textbf{Chemotactic encounters in marine phytoplankton communities.}
The physiological range of exudation rates for healthy phytoplankton cells spans a region crossing the boundary of chemotactic detection ($I_C = 1$, \autoref{fig:ecology}A). 
Up to now, we have considered the excess concentration $C_S$ as a free parameter that could take on a wide range of values. The physiology of marine phytoplankton, however, imposes a coupling between cell size $R$ and excess concentration $C_S$ so that for a phytoplankton of a given size, a narrower range of $C_S$ values is most typical.
Empirical carbon-size scaling laws predict that the carbon content of a phytoplankton cell scales with cell radius as $\sim R^{2.28}$~\cite{mullin1966relationship}. The percent extracellular release (PER), defined as the dissolved fraction of the total primary production~\cite{maranon2005continuity,thornton2014dissolved},  then determines the rate at which the phytoplankton cell leaks carbon to the environment, in the form of dissolved organic matter~\cite{jackson1987simulating,seymour2017zooming} (see SI for details).
PER values are of the order of 10\% (i.e., 10\% of the total primary production is exuded instead of being metabolized for growth), but can range from $2\%$ up to $40\%$, highlighting the strong dependence of the release rate on physiological and environmental conditions~\cite{maranon2004significance}.
Higher PER values may be correlated with higher stress levels of the phytoplankton, resulting in the inability to store or metabolize carbon.
For any given phytoplankton radius $R$, we thus obtain a range of exudation rates (corresponding to a range of PER values) which determine the excess concentration $C_S$ at the surface of the phytoplankton cell (see SI, Equations S1 to S7). Overlaying this range of ecologically relevant values onto the chemotactic performance landscape (\autoref{fig:ecology}A) reveals that chemotaxis might increase encounters across the entire phytoplankton size spectrum. Bacterial detection of the gradients generated by the smallest phytoplankton cells ($R<\SI{3}{\micro\m}$) may only be possible at high PER levels, whereas at low PER levels the gradients may be too tightly confined in space for bacteria to detect them.

The steeply decreasing size structure of marine phytoplankton communities favors encounters of bacteria with small phytoplankton cells.
The average search time $T_e$ required by one bacterium to encounter a phytoplankton cell is~(SI)
\begin{equation}\label{eq:search_time}
    T_e = \dfrac{1}{I_C (R) \Gamma(R) N(R)},
\end{equation}
where $N(R)$ is the concentration of phytoplankton of radius $R$ and $\Gamma(R)$ is the random encounter kernel between bacteria and phytoplankton of size $R$.
Marine phytoplankton communities have size structures that usually follow a power-law size-abundance relationship of the form $N(R) \propto R^{-3\alpha}$, where the allometric exponent $\alpha$ takes values between $\alpha\approx1.0$ in oligotrophic waters and $\alpha\approx0.75$ in productive waters~\cite{cermeno2008species}.
Phytoplankton cells with small radii are, therefore, vastly more abundant than those with large radii, particularly so in oligotrophic environments where large phytoplankton are very rare (\autoref{fig:ecology}B).
Moreover, we consider an overall larger cell abundance in productive waters (a total of \SI{2e5}{cells/\milli\liter} between $0.5$ and \SI{70}{\micro\m} in radius) with respect to oligotrophic waters (a total of \SI{1e5}{cells/\milli\liter} between $0.5$ and \SI{70}{\micro\m} in radius)~\cite{cermeno2008species}. Larger cell abundances are also reflected in higher background chemoattractant concentrations, for which we assume $C_0=\SI{1}{\nano M}$ in oligotrophic waters and $C_0=\SI{100}{\nano M}$ in productive waters~\cite{lee1975amino}.
We first note that, in the absence of chemotaxis, the steep size structure of phytoplankton communities alone results in bacterial search times that are much shorter for small phytoplankton than for larger ones (broken lines in \autoref{fig:ecology}C--D).
In both environments, a faster bacterium with $U=\SI{60}{\micro\m/\s}$ (thin violet broken lines in \autoref{fig:ecology}C and D) will experience shorter search times across the entire size spectrum since faster swimming always increases the rate of random encounters.
We now show that, when chemotaxis is taken into account, the asymmetric performance landscape of bacterial chemotaxis further raises the stakes for bacterial encounters with phytoplankton at the low end of the phytoplankton size spectrum.

Chemotaxis can markedly reduce search times for phytoplankton in the sub-\SI{5}{\micro\m} size range, but different bacterial phenotypes display substantial performance differences in the low end of the phytoplankton size spectrum (\autoref{fig:ecology}C--D).
The decrease in search time afforded by chemotaxis over the whole range of phytoplankton PER values considered is shown as a shaded band in \autoref{fig:ecology}C,D, with the thick marker-decorated line representing $\mathrm{PER}=10\%$.
In oligotrophic waters, slower bacteria ($U=\SI{25}{\micro\m/\s}$) with a longer sensory timescale ($T=\SI{250}{\milli\s}$) decrease their search times for the smallest phytoplankton by almost a factor 10 compared to random motility, whereas faster bacteria ($U=\SI{60}{\micro\m/\s}$) with a shorter sensory timescale ($T=\SI{100}{\milli\s}$) cannot benefit at all from chemotaxis at this low-end of the phytoplankton size spectrum and may be outperformed by the slow swimmers when the phytoplankton PER is high (\autoref{fig:ecology}C).
While the major difference in the $I_C$ values between the two strategies just considered is limited to phytoplankton in the micrometer range, this region of the size spectrum encloses the vast majority of the phytoplankton: in both environments, the 95th percentile of the population (dotted vertical lines in \autoref{fig:ecology}C and D) is below the \SI{2}{\micro\m} radius.
We also note that the boost in chemotactic performance towards small cells occurs only at high PER values, whereas no gain is obtained at intermediate or low PER values. This indicates that chemotaxis towards small phytoplankton may be primarily beneficial when phytoplankton cells are very leaky, as occurs, for example, in the late stages of a bloom or for damaged or senescent individuals~\cite{thornton2014dissolved} (see also SI for discussion on the effect of chemoattractant diffusivity). In contrast, for larger ($>\SI{2}{\micro\m}$) phytoplankton, both phenotypes can benefit from chemotaxis also at lower PER values; while the resulting search times are comparable, the fast swimmers display shorter search times.
In productive waters, where large cells are more abundant and the background concentration of chemoattractants is higher, neither of the two bacterial phenotypes is able to use chemotaxis to reduce search times for phytoplankton cells smaller than $\sim\SI{2}{\micro\m}$ (\autoref{fig:ecology}D). Here the faster bacteria outperform the slower ones in the search for phytoplankton over the entire size spectrum.

To further characterize the impact of chemotaxis on encounters within phytoplankton communities, we integrate the search times (\autoref{fig:ecology}C--D) over the phytoplankton size spectrum. Over the course of \SI{1}{day}, a typical bacterial lifetime~\cite{kirchman2016growth}, an individual slow-swimming bacterium in oligotrophic environments will experience, on average, between 1 random ($\mathrm{PER}=0$) and 13 chemotactic ($\mathrm{PER}=40\%$) encounters, and a faster bacterium will similarly experience 3 random ($\mathrm{PER}=0$) to 12 chemotactic ($\mathrm{PER}=40\%$) encounters. When integrated across the phytoplankton size spectrum, the superior chemotactic performance of slow swimmers towards picophytoplankton thus offsets the random encounter advantage of fast swimmers, supporting the idea that the two search phenotypes may coexist in oligotrophic oceans. In productive waters, a slow swimmer would experience 3 to 14 encounters per day, while a fast swimmer would range between 8 and 21 encounters per day. We can thus conclude that chemotaxis can enhance bacteria-phytoplankton encounters up to ten-fold compared to random motility. To obtain the total number of encounters occuring in a given volume of seawater, these numbers must be multiplied by the concentration of heterotrophic bacteria (\SI{1e5}{cells/\milli\l} in oligotrophic waters and \SI{1e6}{cells/\milli\l} in productive waters, of which we consider only 10\% to be motile)~\cite{mitchell1995long,wigington2016reexamination}. The results predict that there are up to \SI{8e4}{} bacteria-phytoplankton chemotactic encounters occurring every day in a milliliter of oligotrophic waters, and up to \SI{1.7e6}{} in a milliliter of productive waters~(SI, Figure S10).
Moreover, if we assume the individual encounters to be Poissonian events~(SI), the search times will follow an exponential distribution, so that if the average search time is $T_e$, half of the encounters will occur in less than $0.7T_e$, and 10\% of them will occurr in less than $0.1T_e$.

\textbf{Discussion.}
The differences in the chemotactic performance of two representative bacterial phenotypes showcased in \autoref{fig:ecology} highlight how the tradeoff between higher swimming speed and sensing noise in chemotaxis towards small phytoplankton (\autoref{eq:noise_2ndorder}) may only be relevant under certain environmental conditions. Low swimming speeds and long sensory timescales may be considered a high-risk adaptation that provides very high gains: although bacteria with such phenotypes are otherwise outperformed by faster swimmers, they can have superior performance in the search for very leaky picophytoplankton, which may constitute the most favorable niche in conditions where nutrients and large phytoplankton are scarce.
Higher swimming speeds, which provide a higher baseline rate of random encounters at the cost of reduced chemotactic performance in the low end of the phytoplankton size spectrum, are instead a low-risk adaptation, offering robust gains across a wider range of conditions.
We propose that this asymmetry can be a promoter of coexistence between diverse search phenotypes in bacterial populations.

The chemotactic encounter problem can be seen as an evolutionary game where search time is one of the factors influencing the fitness of bacteria.
It is important to keep in mind, however, that the encounter with the sensory horizon is only the first step of a multi-stage interaction process and that fitness is ultimately determined by the ability to grow and reproduce (``growth return'').
A more complete eco-evolutionary picture would require simultaneous knowledge of the search time and the growth return associated with phytoplankton cells of different sizes.
In this respect, chemotaxis is expected to not only reduce search times but also improve growth returns by allowing bacteria to remain near a phytoplankton cell, where concentrations of dissolved organic matter are higher~\cite{fernandez2019foraging}.
Indeed, Raina \textit{et al.}~\cite{raina2023chemotaxis} have shown that \textit{Marinobacter adhaerens} can use chemotaxis to increase both carbon uptake from \textit{Synechococcus} cells and encounter rates with their phycosphere.
Generally, it can be expected that larger phytoplankton, while offering less drastic enhancements in search times through chemotaxis, provide larger growth returns, yet a quantitative consideration of the growth return should also include estimates of the energy expenditure associated with motility (which increases quadratically with swimming speed) and chemotaxis~\cite{malaguti2021theory,keegstra2022ecological}, and the mortality cost associated with predation, which may increase with swimming speed or as a result of extended residence times in nutrient-rich regions~\cite{nielsen2021foraging,ebrahimi2022particle}. Future research in this direction may consider more sophisticated encounter models, able to account for example for hovering, reversible and irreversible attachment, spatial correlations in dense algal blooms, and even destructive effects on the organisms, such as phytoplankton cells being killed by bacteria~\cite{hutchinson2007use,gurarie2013general,mayali2004algicidal,martinez-perez2024space}.

Our results show that the limits of the chemotactic search for small phytoplankton cells are sensitive to the details of motility and sensing. Almost 40 years ago, Jackson hypothesized that the gradients around small cells could not be sensed by a swimming bacterium~\cite{jackson1987simulating}. However, recent observations, such as the aforementioned \textit{Marinobacter adhaerens}--\textit{Synechococcus} interaction~\cite{raina2023chemotaxis} or the investigation of the chemotactic response of marine heterotrophic bacteria to the exudates of virus-infected \textit{Synechococcus} extending to distances larger than \SI{100}{\micro\m}~\cite{henshaw2023early}, indicate that this limit is not universal. Our results provide the foundation that confirms that micrometer-sized targets are generally detectable by chemotactic bacteria and that chemotaxis can significantly reduce the search time for such small cells, under certain conditions~(\autoref{fig:ecology}). Our findings also show that the exact position of the detection boundary is sensitive to the search parameters, including bacterial swimming speed, sensory timescale and chemotactic sensitivity, phytoplankton size and leakage rate, and molecular diffusivity of the chemoattractant~(\autoref{fig:strategies}). Consequently, chemotaxis towards phytoplankton may be more pervasive than previously thought~\cite{seymour2024swimming}, yet questions related to its operational limits or its quantitative benefit for bacterial growth may need to be investigated for each specific system.

More work is needed to evaluate how close to the predicted idealized chemotactic performance bacteria can operate. Agent-based simulations of bacterial chemotaxis towards leaky phytoplankton cells could help to establish the distance to the idealized limit as a function of bacterial phenotypes and chemotactic strategy (i.e., the behavioral response elicited after the detection of a gradient). It is known that different strategies exist~\cite{celani2010bacterial,xie2011bacterial,son2016speeddependent,alirezaeizanjani2020chemotaxis,colin2021multiple}, but the full extent of their diversity and their performance in the context of chemotaxis towards phytoplankton is unclear. Our work indicates that optimal strategies may depend on the phytoplankton cell size --- strategies that work well for large cells will likely underperform for small ones and vice versa. Moreover, experiments designed to assess how far from a phytoplankton cell chemoattractants can be detected by bacteria could aid in providing a stronger connection between the mathematical notion of a sensory horizon (\autoref{fig:sensing}) and the empirically grounded concept of the phycosphere~\cite{seymour2017zooming,bell1972chemotactic,platt2023probing}, the biochemically rich microenvironment which surrounds individual phytoplankton cells.

Beyond bacterial chemotaxis, other forms of taxis and kinesis are widespread in nature, from spatial chemotaxis in leukocytes~\cite{snyderman1981molecular}, through olfactory navigation in insects~\cite{reddy2022olfactory} to visual feeding in fishes~\cite{utne-palm2002visual}. Common features of natural search strategies have been identified~\cite{hein2016natural}, and the predictions of generalized frameworks for sensory searches (e.g., the effect of the search on the stability of antagonistic and mutualistic interactions), depend strongly on the size of the sensory region~\cite{hein2020information}. Our results show that the transition from ballistic to diffusive encounters with the sensory region and the size-dependent limitations on signal perception together impose tradeoffs that determine an asymmetry in the performance of bacterial chemotaxis towards phytoplankton. Due to their fundamental nature, similar tradeoffs may apply to a broader class of systems having different sensory mechanisms.
Investigating mathematical descriptions of the signal-to-noise ratio associated with different sensory systems~\cite{rode2024information} in terms of the search parameters, particularly movement speed, reaction time, and target size, may provide valuable insights into the efficiency and limitations of other natural search strategies.

Our analysis is based on several simplifying assumptions. Phytoplankton exudates typically vary in composition and abundance as a function of the organism's identity, physiological conditions and life stage~\cite{maranon2004significance,thornton2014dissolved,seymour2017zooming}. We only model chemoattractants through their diffusivity, but different bacteria are attracted to distinct compounds with varying sensitivity~\cite{raina2022chemotaxis,mao2003sensitive}, many chemoattractants can drive behavioral and metabolic shifts in bacteria~\cite{egbert2010minimal,ni2020growthrate,barak-gavish2023bacterial,stubbusch2024polysaccharide}, and how the co-occurrence of multiple compounds may impact chemotactic responses is still mostly not understood~\cite{clerc2023strong,li2024phenotypic}. Turbulence in the water column can transport the microorganisms and deform the concentration profiles around phytoplankton cells, enhancing or degrading chemotactic performance in ways that strongly depend on turbulent intensity, phytoplankton size and bacterial speed~\cite{bowen1993simulating,visser1998turbulenceinduced,taylor2012tradeoffs,lange2021sperm}, although at the low intensities typical of the ocean~\cite{franks2022oceanic} we may expect its contribution to mostly enhance encounters.
Temporal fluctuations, from daily to seasonal cycles, drive variations in cell abundances~\cite{benedetti2019seasonal,haentjens2022phytoplankton}, rescaling the search times proportionally, as well as in leakage rates~\cite{prezelin1992diel} and bulk concentrations of chemicals~\cite{takahashi1993seasonal}. All these factors and others, such as predation~\cite{nielsen2021foraging}, sinking~\cite{kiorboe2001marine,slomka2020encounter} and viral infections~\cite{zimmerman2020metabolic,delong2023consumption}, contribute to shaping the microscale interaction landscape of the ocean.

\textbf{Conclusions.}
Using idealized models of phytoplankton leakage and bacterial chemotaxis, we calculated upper bounds on the enhancement in bacteria-phytoplankton encounters driven by chemotaxis over random motility alone and studied how the enhancement depends on the size of the phytoplankton and on bacterial phenotypes. We found that bacterial chemotaxis offers low-risk/low-gain performance in the search for large phytoplankton but high-risk/high-gain performance in the search for small phytoplankton. This fundamental tradeoff arises from chemotactic encounters being limited by different mechanisms at the two ends of the phytoplankton size spectrum. For large phytoplankton, the limitation stems from the small steepness of the gradients and the diffusive nature of bacterial motility, which leads to moderate but consistent reductions in search times due to chemotaxis. By contrast, for small phytoplankton, the limitation stems from the small spatial extent of the gradients and the ballistic motility of bacteria, which leads to large yet less reliable reductions in search times due to chemotaxis. Furthermore, we found that searching for small phytoplankton is more efficient when bacteria swim more slowly and integrate the gradients over longer sensory timescales. Overall, our results suggest that the high-stakes nature of encounters with small phytoplankton is a fundamental feature of chemotactic searches that may drive a diversity of size-sensitive chemotactic strategies among marine bacteria.

\textbf{Materials and methods.} All the data shown in this work has been generated through custom Julia (version 1.10) code, which is available on GitHub: \url{https://github.com/mastrof/GradientSensing}.
Project codes and data were managed with DrWatson.jl~\cite{datseris2020drwatson}.
Numerical solution of the equations for the evaluation of the sensory horizon $S$ was performed with the \texttt{Order0} method~\cite{kahan1979personal} implemented in the Roots.jl library.
Simulations for the Kolmogorov-Smirnov model sensor made use of Distributions.jl~\cite{dahualin2024juliastats} and HypothesisTests.jl.
Figures were realized with Makie.jl~\cite{danisch2021makie} and Inkscape. Interactive dashboards were realized with Makie.jl~\cite{danisch2021makie}.

\textbf{Acknowledgements.}
We thank Johannes Keegstra and Thomas Ki{\o}rboe for discussions. We gratefully acknowledge funding from the European Union’s Horizon 2020 research and innovation program under Marie Sk\l{}odowska-Curie grant no. 955910 to R.F.; the \lq Agence Nationale de la Recherche\rq~under grant ANR-22-CPJ2-0015-01 to F.J.P.; 
a Gordon and Betty Moore Foundation Symbiosis in Aquatic Systems Initiative Investigator Award (\href{https://doi.org/10.37807/GBMF9197}{GBMF9197}), the Simons Foundation through the Principles of Microbial Ecosystems (PriME) collaboration (grant 542395FY22), Swiss National Science Foundation grant \texttt{205321\char`_207488}, Swiss National Science Foundation Sinergia grant CRSII5-186422, and the Swiss National Science Foundation, National Centre of Competence in Research (NCCR) Microbiomes (No. \texttt{51NF40\char`_180575}) to R.S.;
a Swiss National Science Foundation Ambizione grant no. \texttt{PZ00P2\char`_202188} to J.S.
We also gratefully acknowledge ETH Z\"urich (Euler cluster) for providing computational resources.

\section{Supplementary Material}
\renewcommand{\theequation}{S\arabic{equation}}
\setcounter{equation}{0}

\subsection*{Concentration fields generated by phytoplankton cells}
In this section, we describe in more detail the parameterization of the concentration fields around phytoplankton cells used in the Main Text.

At any stage in their life cycle, phytoplankton cells exude a variety of organic compounds in the water columnn~\cite{thornton2014dissolved}, many of which can be strong bacterial chemoattractants~\cite{raina2022chemotaxis}. These compounds give rise to a biochemically rich microenvironment known as the phycosphere~\cite{seymour2017zooming}, where microscale interactions with other organisms take place.
A clear physico-chemical description of the phycosphere is still missing, but some empirical models have been developed to provide a simple qualitative picture. 

In his 1987 study, Jackson proposed a model of phytoplankton leakage based on the total carbon content of a cell~\cite{jackson1987simulating}. It has been observed that different phylogenetic groups show similar carbon-volume relationships, with the cell carbon content scaling as a power-law of the cell volume. Although more modern measurements exist~\cite{menden-deuer2000carbon}, where a difference is also highlighted between diatoms and other protists, for consistency with other works~\cite{jackson1987simulating,seymour2017zooming} we adopt the results from Mullin \textit{et al.}~\cite{mullin1966relationship}, where it is estimated that the carbon content of a cell of volume $V$ follows $\log_{10}\mathrm{C} = 0.76\log_{10}V - 0.29$ where $\mathrm{C}$ is in units of picograms and $V$ in units of \si{\micro\m^3}.
Assuming a spherical shape for the phytoplankton cell and converting the mass to a molar concentration, this carbon-volume relationship is reformulated as
\begin{equation}
    \phi = b R^{2.28}
\end{equation}
where $\phi$ is the molar concentration of carbon in the cell, and $b=\SI{1.67e-4}{\mol/\centi\m^{2.28}}$ is an empirical constant.

If all this carbon was channeled into a single compound, containing $n_C$ carbon atoms per molecule, then the molar concentration of this compound inside the cell would be $\phi/n_C$.
The rate of production of this compound would then be equal to the product of the cell growth rate ($\mu$), and its total abundance, $\phi/n_C$; if we then assume that a certain fraction of this production rate is exuded to the surrounding medium, as described by the percent extracellular release (PER), we can write the exudation rate of a phytoplankton cell as
\begin{equation}\label{eq:leakage_carbon}
    L = \dfrac{\phi}{n_C}\mu \mathrm{PER}.
\end{equation}

If the exuded compound diffuses out of the cell into the water, we can write the dynamics of the system as a diffusion equation
\begin{equation}\label{eq:diffusion_equation}
    \pdv{C}{t} = D\nabla^2 C
\end{equation}
with a boundary condition forcing the concentration far away from the cell to converge towards the bulk value of the concentration ($C_0$)
\begin{equation}
    \lim_{r\to\infty} C(r) = C_0
\end{equation}
and another boundary condition to account for the exudation with rate $L$
\begin{equation}
    \lim_{r\to0} 4\pi D_C r^2\dv{C}{r} = -L
\end{equation}
we obtain, at steady-state, the concentration field
\begin{equation}\label{eq:concentration}
    C(r) = C_0 + C_S\dfrac{R}{r}, \qquad r \geq R
\end{equation}
where
\begin{equation}\label{eq:leakage}
    C_S = \dfrac{L}{4\pi D_C R}
\end{equation}
is the concentration at the phytoplankton cell surface in excess to the background concentration, and $r$ is the radial distance from the center of the spherical phytoplankton cell.

\subsection*{Small phytoplankton cells generate steeper gradients}
The gradient associated with the concentration field of \autoref{eq:concentration} is
\begin{equation}
    \nabla C(r) = -C_S\dfrac{R}{r^2}\hat{\mathbf{r}}, \qquad r \geq R.
\end{equation}
To illustrate why, for a fixed excess concentration $C_S$, the gradients generated by small phytoplankton cells are steeper, and therefore more tightly confined in space, we express the gradient in terms of the distance from the surface of the phytoplankton ($\delta$), rather than from its center: $r = R + \delta$ with $\delta > 0$. 
We also divide the gradient by $C_S$ to scale out the dependence on the excess concentration, which yields the steepness
\begin{equation}
    s \equiv \dfrac{|\nabla C(\delta)|}{C_S} = \dfrac{R}{(R+\delta)^2} = \dfrac{1}{R}\dfrac{1}{(1+\delta/R)^2}.
\end{equation}
Close to the surface of the phytoplankton cell ($\delta\ll R$) the steepness can be approximated as
\begin{equation}\label{eq:gradient_steepness}
    s \approx \dfrac{1}{R}\qty(1 - \dfrac{2\delta}{R})
\end{equation}
which is larger for smaller values of the phytoplankton radius $R$. As we move far from the cell ($\delta\gg R$) we have instead
\begin{equation}
    s \approx \dfrac{R}{\delta^2}
\end{equation}
which increases proportionally to the phytoplankton radius $R$.
These two limits highlight that the concentration fields generated by smaller phytoplankton cells are steeper in the proximity of the cell  compared to the proximity of large cells, but, as a consequence, they decay much faster to the background, and therefore at larger distances they are less steep than those generated by larger cells.

\subsection*{On the validity of the steady-state assumption}
In the calculation above, and in all of the text, we assume concentration profiles around phytoplankton cells to be at steady state. Here, by solving the time-dependent diffusion equation, we show that this assumption is warranted because the timescale of the relaxation to steady state is much shorter than the typical timescales involved in chemotactic searches.

In spherical coordinates, and with the boundary conditions defined above, the complete time-dependent solution of \autoref{eq:diffusion_equation} is~\cite{crank1975mathematics}
\begin{equation}\label{eq:time-dependent-diffusion}
    C(r,t) = C_0 + C_S\dfrac{R}{r}\tn{erfc}\qty(\dfrac{r-R}{2\sqrt{D_C t}})
\end{equation}
where $\tn{erfc}$ is the complementary error function.
The characteristic timescale for molecules diffusing from the phytoplankton surface to a distance $r$ can then be defined as
\begin{equation}\label{eq:diffusive-timescale}
    \tau(r) = \dfrac{(r-R)^2}{4 D_c}.
\end{equation}

For a diffusivity $D_C=\SI{500}{\micro\m^2\per\s}$, the relaxation timescale for distances up to $r-R=\SI{100}{\micro\m}$ is below $\SI{5}{\s}$. The decay of the time-dependent term is, however, relatively slow, especially when compared to typical exponential relaxation processes. It is therefore reasonable to expect that, for the time-dependent solution to become ``sufficiently'' close to the steady state, the time required will be a multiple of the timescale $\tau$.
We may consider the solution as ``converged'' when it reaches 90\% of its steady state value. This equates to equilibration times $\sim125\tau(r)$. \autoref{fig:time-dependent-diffusion} shows the ratio between the time-dependent solution of the diffusion equation (\autoref{eq:time-dependent-diffusion}) and the steady state solution (\autoref{eq:concentration}) as a function of both time, $t$, and radial distance from phytoplankton surface, $r-R$. The boundary layer up to $\sim\SI{20}{\micro\m}$ from the surface relaxes to steady state within $\sim\SI{10}{s}$, and it takes less than 10 minutes for the relaxation at distances up to \SI{100}{\micro\m}.

Since the timescales of chemotactic encounters are much longer (as shown in Fig.~5 in the main text), it is reasonable to assume the diffusive concentration profile around phytoplankton cells to be instantaneously at steady state from the point of view of chemotactic bacteria.

\subsection*{Diffusion with consumption in the bulk: screened decay}
While a purely diffusive profile might be realistic around a cell in isolation, we can expect that within marine environments there will be various factors that perturb the concentration profile. A relevant scenario to consider is bacterial consumption of the exuded compound.

A typical concentration of bacteria in seawater is $B\sim10^6\,\si{bacteria\per\milli\liter}$~\cite{wigington2016reexamination}. If the bacteria consume the compound exuded by the phytoplankton cell, this will affect the concentration profile of the compound.
Assuming the uptake to be diffusion-limited and the bacteria to be uniformly distributed in the medium, the steady-state equation for the concentration profile can be written as
\begin{equation}\label{eq:diffusion_sink}
    D_C\nabla^2C = 4\pi D_C B aC
\end{equation}
where $a$ is the radius of bacteria and $4\pi D_C B a$ is therefore the rate at which the compound is being consumed.
With the same boundary conditions as the previous case, the solution to this equation is the concentration field
\begin{equation}\label{eq:expfield}
    C(r) = C_0 + C_S\dfrac{R}{r}\exp(-\dfrac{r-R}{\rho}), \qquad r\geq R
\end{equation}
where $C_S = L/(4\pi D_C R)$ is the excess concentration at the phytoplankton surface, and $\rho = 1 / \sqrt{4\pi aB}$, a quantity with units of length, represents the characteristic lengthscale for the exponential decay of the concentration profile. That is, the presence of a bacterial population consuming the compound has a screening effect on the concentration field which decays to the bulk much faster than the ideally diffusive field, with the characteristic decay lengthscale set by $\rho$.
Notice that in the limit $\rho \to \infty$ (corresponding to $B\to 0$, i.e. infinite bacterial dilution and therefore no consumption), this result converges to the purely diffusive concentration field (\autoref{eq:concentration}) used in the main text and discussed in the previous section.
The shape of the screened concentration field (\autoref{eq:expfield}) for some representative values of the screening length $\rho$, as well as the limit case $\rho\to\infty$ corresponding to \autoref{eq:concentration}, are shown in \autoref{fig:concentration-fields}

This concentration profile shows no qualitative differences from the ideal diffusion profile considered in the main text. Minor quantitative differences, resulting from the fact that for any given phytoplankton radius the screened concentration profile will decay faster than the corresponding ideal diffusion profile, do not affect our conclusions.
To quantify the effect of consumption on chemotactic performance, we estimate the chemotactic performance landscape exactly as done for the ideal diffusion profile (\autoref{fig:ic_expfield}). We notice that only for small values $\rho=\SI{30}{\micro\m}$, corresponding to an extremely high bacterial concentration $B\sim\SI{1.8e8}{cells/\milli\liter}$, the landscape deviates from that evaluated for the ideal diffusion profile (Figure~3 in the main text), but the deviation is small and only affects the larger cells which, due to the gradients being steeper, can be detected at lower excess concentration $C_S$ (\autoref{fig:ic_expfield}A).
For smaller bacterial concentrations (and thus longer screening lengths $\rho$) the effect is negligible (\autoref{fig:ic_expfield}B-C).

Analyzing the gradient of this concentration field shows no qualitative difference from the ideal diffusion case. The steepness here takes the form
\begin{equation}
    s \equiv \dfrac{|\nabla C(r)|}{C_S} = \dfrac{R}{r}\dfrac{\rho + r}{\rho r}
    \exp(-\dfrac{r-R}{\rho}).
\end{equation}
As $r\to R$, we have
\begin{equation}
    s \approx \dfrac{\rho+R}{\rho R}\qty(1 - \dfrac{r-R}{\rho}).
\end{equation}
If $\rho$ is large with respect to the phytoplankton radius $R$, uptake has a small effect on the concentration field and the leading term to the gradient steepness is again $s\sim1/R$ as in \autoref{eq:gradient_steepness}.
When $\rho$ is small instead, i.e. there is strong consumption, the leading term in the steepness close to the cell surface becomes $1/\rho$ which is independent of phytoplankton size. With reasonable bacterial concentrations, however, $\rho$ is always large enough to have nearly no effect in close proximity to the cells: e.g., for a bacterial concentration $B=10^6\,\si{cells\per\milli\liter}$ we have $\rho\approx\SI{400}{\micro\m}$, and with $B=10^7\,\si{cells\per\milli\liter}$ it only goes down to $\rho\approx\SI{126}{\micro\m}$.
Even with consumption effects, small phytoplankton cells generate concentration fields that are steeper and more tightly confined in space.

To deepen the understanding of the consequences of bacterial consumption, we also consider more extreme screening, e.g. $\rho=\SI{3}{\micro\m}$. For diffusion-limited consumption by bacteria with \SI{0.5}{\micro\m} radius, the bacterial concentration required to achieve this value would be $B\approx\SI{1.8e10}{cells/\milli\l}$. This is a very high concentration for typical oceanic conditions. However, strong screening may also occur due to consumption by larger organisms, in addition to other effects which may contribute as sink terms to \autoref{eq:diffusion_sink}, e.g. thermal or chemical degradation of the compound.
In such an extreme scenario, we observe a noticeable deviation from the ideal case. The fundamental features of the chemotactic performance landscape are, however, conserved (\autoref{fig:extreme-screening}). In general, the boundary of detection is pushed towards lower $C_S$ values, because the concentration profile decays over a very short distance, generating very steep gradients. For small phytoplankton, we still observe the sharp transition in $I_C$ as the boundary of detection is crossed. Notably, for large phytoplankton, the transition across the boundary of detection becomes even smoother. Since the screening is so strong, a large increase in $C_S$ is required to produce a tiny expansion of the sensory horizon and a consequent increase in $I_C$.

\section*{Dynamic correction of the sensing noise}
Fluctuations in molecular adsorption events lead to noise in bacterial estimates of the local concentration gradient~\cite{berg1977physics}. In the case of linear concentration fields (i.e., constant gradients), Mora \& Wingreen~\cite{mora2010limits} have evaluated the theoretical minimum noise affecting the gradient measurement:
\begin{equation}\label{eq:noise_mw}
    \sigma_0(r) = \sqrt{\dfrac{3C(r)}{\pi a D_C T^3}},
\end{equation}
where $a$ is the radius of the bacterium (assumed to be spherical), $D_C$ is the diffusivity of the chemoattractant and $T$ is the sensory timescale of the bacterium.

In the scenario considered by Mora and Wingreen, the swimming speed of the bacterium has no effect on the sensing noise. However, the assumption of constant gradients is not generally valid for the non-linear, spatially confined concentration fields around phytoplankton cells. In what follows, we follow the same approach of Mora and Wingreen to show that the presence of non-linearities in the concentration fields introduces speed- and position-dependent (``dynamic'') contributions to the sensing noise.

As a bacterium swims, it is exposed to a chemoattractant concentration $C(t)$ that varies in time ($t$). In a linear concentration field, this time-dependent concentration would be expressed as $C(t) = c_0 + c_1 t$, and the task of gradient detection would correspond to the estimation of the ramp rate $c_1$. This is what Mora and Wingreen considered: the sensing noise $\sigma_0$ (\autoref{eq:noise_mw}) is, therefore, the standard deviation associated with bacterial estimates of $c_1$.
Here, we consider a non-linear field $C(t) = c_0 + c_1 t + c_2 t^2$.
If, over the sensory interval $[-T/2, +T/2]$, the bacterium is exposed to a timeseries of absorption events
\begin{equation}
    I(t) = \sum_{i=1}^n \delta(t-t_i)
\end{equation}
where $\delta$ is the Dirac delta, $n$ is the total number of absorption events that occured within the interval, and $t_i$ are the times at which the events occurred, an estimate of the field can be obtained by performing a linear regression of this timeseries to the theoretical expected rate of absorption events, $4\pi D_C a C(t)$, where $D_C$ is the molecular diffusivity of the chemoattractant and $a$ the radius of the bacterium.
The linear regression problem then provides the estimates $\hat c_0$, $\hat c_1$ and $\hat c_2$ for the three parameters that define the concentration field. Each estimate is obtained by solving
\begin{equation}
    \pdv{c_n}\int_{-T/2}^{+T/2}\dd{t}\qty[
        \sum_{i=1}^n\delta(t-t_i) - 4\pi D a C(t)
    ]^2 = 0
\end{equation}
which can be more conveniently reformulated as
\begin{equation}\label{eq:linear_regression}
    \int_{-T/2}^{+T/2}\dd{t}\qty[
        \qty( 4\pi D a C(t) - \sum_{i=1}^n\delta(t-t_i) )\pdv{C(t)}{c_n}
    ] = 0.
\end{equation}
The optimal estimates for the parameters are then
\begin{align}
    \hat c_0 &= \dfrac{9}{4}\dfrac{n}{4\pi D a T}\qty[ 1 - 15\dfrac{\sum_i (t_i/T)^2}{n} ] \\[8pt]
    \hat c_1 &= \dfrac{12\sum_i t_i}{4\pi D a T^3} \\[8 pt]
    \hat c_2 &= \dfrac{15\qty[12\sum_i (t_i/T)^2] - n}{4\pi D a T^3}.
\end{align}
Since the first derivative of the field is linear, $c'(t) = c_1 + 2c_2t$, the average gradient experienced by the bacterium throughout the $[-T/2,+T/2]$ interval is $c_1$. Therefore, we assume that the gradient measurement is equivalent to estimating the coefficient $c_1$, as it was in the linear ramp case originally considered by Mora and Wingreen (where $c_2=0$).
The difference here is that the noise affecting the estimate of $c_1$ will also depend on the non-linear term $c_2$, that is, on the curvature of the field.
The variance in the $\hat c_1$ estimate can be found as
\begin{equation}
    \expval{\delta\hat c_1 ^2} = \qty(\dfrac{12}{4\pi D a T^3})^2 \expval{\delta\qty(\sum_{i=1}^n t_i)^2}
\end{equation}
where the $\delta$ in the angular brackets represents the statistical fluctuation, i.e., $\expval{\delta x^2} = \expval{x^2}-\expval{x}^2$.
Since the absorption events are independent, we obtain
\begin{equation}
    \expval{\delta\hat c_1 ^2} = \qty(\dfrac{12}{4\pi D a T^3})^2
    \qty[ 4\pi D a \qty( c_0\dfrac{T^3}{12} + c_2\dfrac{T^5}{80} ) ] = 
    \dfrac{12}{4\pi D a T^3}\qty(c_0 + \dfrac{3}{20}c_2T^2)
\end{equation}
and hence
\begin{equation}\label{eq:sensing_noise_2ndorder}
    \sigma_1  = \sqrt{\expval{\delta\hat c_1^2}} = \sqrt{\dfrac{3}{\pi D a T^3}\qty(c_0 + \dfrac{3}{20}c_2T^2)}
    = \sigma_0\sqrt{1 + \dfrac{3}{20}\dfrac{c_2T^2}{c_0}}.
\end{equation}
Therefore, the non-linear term in the concentration field introduces a correction to \autoref{eq:noise_mw} which increases (since we take $c_2>0$) the gradient sensing noise.

Now we can express the coefficients $c_0$, $c_1$, and $c_2$ in terms of the coefficients of the second order expansion of the diffusive $1/r$ field. If the radial position of the bacterium a the time $t=0$ is $\bar{r}$ and it moves radially towards the source at speed $U$, then
\begin{equation}\label{eq:field_2ndorder_expansion}
    C_S\dfrac{R}{\bar{r}-Ut} \approx C_S\dfrac{R}{\bar{r}}\qty(1 + \dfrac{Ut}{\bar{r}} + \qty(\dfrac{Ut}{\bar{r}})^2),
\end{equation}
from which we identify the coefficients of the time-dependent $C(t)$ as
\begin{align}
    c_0 &= C_S\dfrac{R}{\bar r}, \\
    c_1 &= C_S\dfrac{R}{\bar r}\dfrac{U}{\bar r},\\
    c_2 &= C_S\dfrac{R}{\bar r}\dfrac{U^2}{\bar r^2}.
\end{align}
The correction term to the sensing noise, \eqref{eq:sensing_noise_2ndorder}, then becomes
\begin{equation}
    \xi \equiv \sqrt{1 + \dfrac{3}{20}\dfrac{c_2T^2}{c_0}} = \sqrt{1 + \dfrac{3}{20}\qty(\dfrac{UT}{\bar r})^2}.
\end{equation}
We can further express the radial position of the bacterium in terms of its radial distance from the surface of the chemoattractant source, $\bar r = R + \Delta r$ so that
\begin{equation}\label{eq:noise_correction}
    \xi = \sqrt{1 + \dfrac{3}{20}\qty(\dfrac{UT}{R+\Delta r})^2}.
\end{equation}
\eqref{eq:noise_correction} implies that:
\begin{itemize}
    \item as the swimming speed, $U$, or the sensory timescale, $T$, are increased, the noise in the gradient estimate increases;
    \item as the distance of the bacterium from the surface of the source, $\Delta r$, decreases, the noise in the gradient estimate increases;
    \item the correction term resulting from the non-linearity only becomes relevant when $R+\Delta r$ is significantly smaller than $UT$, which can only happen when the bacterium is close to the source \emph{and} the source is small relative to the physical distance traveled during a single sensory interval.
\end{itemize}

In \autoref{fig:ic_noise}, we show the impact of the dynamic correction to the sensing noise (\autoref{eq:noise_correction}) on the chemotactic performance landscape. We first consider the result for constant gradients from Mora and Wingreen (\eqref{eq:noise_mw}), which assumes a large spatial extent of the gradients. To this end, we solve Equation~3 of the main text numerically for $S$ with $f=1$ (i.e., there is no low-pass filter) and use the solution to compute $I_C$ according to Equation~4 of the main text to plot the chemotactic performance landscape. In this case, we observe that, as the phytoplankton radius decreases, the minimal excess concentration required to elicit chemotactic detection keeps decreasing (\autoref{fig:ic_noise}A). Although the associated chemotactic index would be small, this result would imply that, in principle, any source size could be detected, a statement that in general cannot hold true once the final spatial extent of the gradient is taken into account.
When the dynamic correction to the sensing noise (\autoref{eq:noise_correction}) is included, we observe that the boundary of detection plateaus at low values of $R$: when the radius of the phytoplankton $R$ becomes smaller than the distance traveled by the bacterium during the sensory timescale, $UT$, the increased steepness of the gradients generated by small phytoplanktons is counterbalanced by their spatial confinement (\autoref{fig:ic_noise}B).

The second order expansion (\autoref{eq:field_2ndorder_expansion}) is clearly insufficient to account for the full nonlinearity of the concentration field (\autoref{eq:concentration}), especially in the proximity of small sources. This is evident in the shape of the performance landscape, which exhibits the plateau for small phytoplankton size (\autoref{fig:ic_noise}B) rather than the U-shape (Fig.~3 of the main text). Higher order terms, however, cannot be dealt with in such a simple manner: the estimate of the ramp rate $\hat c_1$ will not be enough to define the average gradient experienced by the bacterium during the sensory interval, which will instead depend on all the odd coefficients in the series expansion ($\hat c_1, \hat c_3, \ldots$). The linear regression problem associated to a higher order expansion can be solved using computer algebra systems, but the resulting uncertainty in the coefficient estimates will contain multiple cross-correlation terms whose calculation may not be feasible.
We thus rely on qualitative arguments to suggest that stronger non-linearity (and hence further terms in the series expansion of the concentration field) must be associated with an increase in the sensing noise and ultimately leads to the convex performance landscape discussed in the main text. First, reason suggests that as the radius $R$ of the source decreases towards 0, the required excess concentration must dramatically increase in order for the source to be detectable; it is unphysical to expect that the minimal $C_S$ required to detect a source plateaus after a critical value of $R$ (as seen in \autoref{fig:ic_noise}B).
Second, the nature of the expansion (\autoref{eq:field_2ndorder_expansion}) shows that stronger non-linearities are associated with higher powers of the $UT/R$ ratio, powers that (despite the cross-correlations) will also appear in the sensing noise. It is therefore reasonable to expect that an order-$K$ expansion of the concentration field will lead to the appearance of a leading term $\mathcal{O}((UT/R)^K)$ in the sensing noise. Ultimately, these considerations are supported by the Kolmogorov-Smirnov sensor, which naturally exhibits the convex performance landscape (Figure 4D--F of the main text) without any assumption on the sensing noise.

\section*{Evaluation of the sensory horizon}
In the main text (Equation 3), we defined the sensory horizon in terms of the signal-to-noise ratio (SNR) experienced by a bacterium measuring the local chemoattractant gradient:
\begin{equation}\label{eq:snr}
    \SNR = \dfrac{|U\nabla C(S)|f(UT/S)}{\Pi\sigma_0(S)} = q,
\end{equation}
where $f$ is the lowpass filter associated with the dynamic component of sensing noise, that we empirically choose as $f(x) = 1 - \exp(-x^{3/2})$, $\sigma_0$ is the inherent sensing noise as defined by Mora and Wingreen~(\autoref{eq:noise_mw}), $\Pi$ is the chemotactic precision factor which accounts for other sources of noise internal to chemotaxis pathway and that was estimated as $\Pi\approx 6$ for marine bacterium \textit{Vibrio anguillarum}, and $q$ is a threshold on the value of the $\SNR$ required for gradient detection.
In particular, for the concentration field in \autoref{eq:concentration}, this relationship reads
\begin{equation}\label{eq:snr_explicit}
    \SNR = \dfrac{1}{\Pi}\dfrac{UT}{S}\sqrt{\dfrac{\pi a D_C T C_S}{3}\dfrac{R}{S}}f(UT/S) = q.
\end{equation}
To evaluate the size of the sensory horizon we solve this equation for $S$ numerically using the `Order0` solver~\cite{kahan1979personal} from the `Roots.jl` Julia package.

Of particular importance is the choice of the threshold value $q$, which determines the minimal $\SNR$ required by a bacterium to robustly discern a gradient from the background noise.
A single measurement of the gradient can be represented as a Gaussian process with mean $\mu>0$ and standard deviation $\sigma>0$, with each position in space being represented by different values of $\mu$ and $\sigma$. The probability to measure a positive gradient at a given position in space is therefore given by the probability to extract a positive value from a given normal distribution
\begin{equation}
    p(\nabla > 0) = \int_0^{+\infty}\dd{x} \mathcal{N}(x; \mu,\sigma) =
    \dfrac{1}{\sqrt{2\pi}\sigma}\int_0^{+\infty}\dd{x} \exp(-\dfrac{(x-\mu)^2}{2\sigma^2}) =
    \dfrac{1}{2}\qty(1 + \mathrm{erf}\qty(\dfrac{\SNR}{\sqrt{2}})).
\end{equation}
where we identified the signal-to-noise ratio with $\SNR = \mu/\sigma$ and $\mathrm{erf}$ is the error function. At any position in space, the SNR determines the probability that a bacterium performs a positive measurement of the gradient: as $\SNR\to\infty$, the probability to correctly detect a positive gradient goes to 1; when $\SNR\to0$, this probability converges instead to 1/2, with each measurement becoming equivalent to a coin flip.
By setting a threshold value on the SNR, we can therefore identify at what distance from the phytoplankton cell the bacteria have a certain probability to measure a positive gradient.
This choice is clearly arbitrary and can lead to more or less relevant differences in the quantitative outcomes of our analysis, but cannot qualitatively alter our conclusions.
Choosing $q=1$ as the threshold for sensing, as we do in the main text, is equivalent to requesting that $p(\nabla>0)\approx0.841\ldots$.
With this convention, the sensory horizon $S$ is therefore the distance at which a bacterium heading towards the phytoplankton cell is able to measure a positive gradient roughly 85\% of the time.

\section*{Evaluation of the chemotactic index}
To determine the performance of a chemotactic search, we introduced the chemotactic index $I_C$ as the ratio between the rate of chemotactic encounters and the rate of random encounters.
Encounter rates are classically measured in terms of encounter kernels, physical quantities which represent the volume that a pair of bodies sweep relative to each other per unit time~\cite{slomka2023encounter}.
Given two species $A$ and $B$, with concentrations $c_A$ and $c_B$ respectively, then
\begin{equation}
    \text{encounter rate per unit volume} = \Gamma_0 c_A c_B
\end{equation}
where $\Gamma_0$, the random encounter kernel, contains all the information regarding the relative motion of $A$ and $B$.

Consider a bacterium performing a random walk with correlation length $\lambda$ and speed $U$; we will assume the bacterium to be spherical with radius $a$.
The phytoplankton cell, also spherical, with radius $R$ is assumed to be stationary.
The encounter kernel between the randomly moving bacterium and the phytoplankton cell is determined by the relationship between the size of the two organisms ($a$ and $R$) and the correlation length of the bacterial random walk ($\lambda$). Despite the apparent simplicity of the system, an analytical formula for the encounter kernel is available only in two limit scenarios.

When $\lambda \ll a+R$ the encounters are diffusion-limited and are thus described by the kernel first introduced by Smoluchowski in the context of coagulation theory~\cite{chandrasekhar1943stochastic}:
\begin{equation}\label{eq:kernel_diffusive}
    \Gamma_0 = 4\pi D (R+a)
\end{equation}
where $D$ is the effective diffusivity of the bacterium, which can be expressed as $D=U\lambda/3$~\cite{lovely1975statistical,berg1993random}.
Note that we are explicitly talking about ``correlation length'' of the bacterial motion rather than ``run length'': this allows us to assume that any correction factor associated to angular correlations or different motility patterns is already contained within $\lambda$.

In the opposite limit, where $\lambda \gg a+R$, the nature of encounters is instead ballistic and is described by Maxwell's kinetic theory of gases~\cite{maxwell1860illustrations}. The corresponding encounter kernel is
\begin{equation}\label{eq:kernel_ballistic}
    \Gamma_0 = \pi U (R+a)^2.
\end{equation}

Outside these two limits, in the so-called transition regime, analytical descriptions of the encounter process are not available and it is necessary to recur to empirical formulas.
Accurate empirical expressions covering the three regimes can be found in the literature~\cite{gopalakrishnan2011determination}, but for this study we decided to adopt the simple (although, clearly, less accurate) interpolation formula~\cite{visser2007motility}
\begin{equation}
    \Gamma_0 = \dfrac{\mathrm{Kn}}{1 + \mathrm{Kn}} \pi U (R+a)^2
\end{equation}
where we defined the Knudsen number
\begin{equation}
    \mathrm{Kn} = \dfrac{4\lambda}{3(R+a)}
\end{equation}
which can be rationalized as a measure of the ``ballisticity'' of the encounter.
$\mathrm{Kn}\ll1$ represents the diffusive regime, where we recover \eqref{eq:kernel_diffusive}, while $\mathrm{Kn}\gg1$ represents the ballistic regime, where we recover \eqref{eq:kernel_ballistic}.

While performing calculations, we will mantain the dependency on both $R$ and $a$; for our particular system, however, we can fix $a\sim\SI{0.5}{\micro\m}$ and assume that the condition $\lambda\gg a$ is always satisfied, as motile bacteria display correlation lengths much larger than their body size. This implies that, whenever $\lambda \ll R+a$, then it must be that $R\gg a$ and therefore we can ignore the size of bacteria, giving $R+a \approx R$.

In the perfect chemotaxis framework, we modeled chemotactic encounter rates as random encounters with a sensory horizon of radius $S$. In this case, the framework of random encounter rates can be immediately applied by simply replacing the real phytoplankton cell radius $R$ with $S$. Defining a sensory amplification factor
\begin{equation}
    \eta = \dfrac{S+a}{R+a} \geq 1
\end{equation}
which quantifies the increase in the apparent (total) size of the encountering organisms as a result of chemotaxis, we can then express the chemotactic encounter kernel as
\begin{equation}
    \Gamma = \dfrac{\eta^2\mathrm{Kn}}{\eta+\mathrm{Kn}}\pi U (R+a)^2.
\end{equation}
Therefore, the chemotactic index takes the form
\begin{equation}\label{eq:ic_general}
    I_C = \dfrac{\Gamma}{\Gamma_0} = \dfrac{1+\mathrm{Kn}}{\eta+\mathrm{Kn}}\eta^2.
\end{equation}
This is the equation we use throughout this work to determine the $I_C$ values.
In the diffusive limit $\mathrm{Kn}\ll1$, the chemotactic index becomes
\begin{equation}
    I_C = \eta = \dfrac{S+a}{R+a} \approx \dfrac{S}{R}, \quad (\mathrm{Kn} \ll 1).
\end{equation}
In the ballistic limit $\mathrm{Kn}\gg1$, the chemotactic index becomes
\begin{equation}
    I_C = \dfrac{\mathrm{Kn}}{\eta+\mathrm{Kn}}\eta^2,
\end{equation}
which can still give rise to two possible scenarios, depending on the relationship between $\eta$ and $\mathrm{Kn}$ (or, in other words, between $S+a$ and $\lambda$).
If $\eta \ll \mathrm{Kn}$, we have
\begin{equation}
    I_C = \eta^2 = \qty(\dfrac{S+a}{R+a})^2, \quad (\mathrm{Kn} \gg \eta \gg 1).
\end{equation}
This case, corresponding to ballistic encounters in which the sensory amplification is small, implies that the sensory horizon $S$ is small enough that random encounters with it still retain a ballistic nature ($4\lambda/3(S+a) \ll 1$).
In the $\eta \gg \mathrm{Kn}$ case, we find instead
\begin{equation}
    I_C = \eta\mathrm{Kn} = \dfrac{4\lambda(S+a)}{(R+a)^2} \approx \dfrac{4\lambda S}{(R+a)^2}, \quad (\eta \gg \mathrm{Kn} \gg 1).
\end{equation}
This is a mixed regime where random encounters with the phytoplankton cell are ballistic, but the sensory amplification is so strong that random encounters with the sensory horizon display a diffusive rather than ballistic nature ($S+a\gg\lambda\gg R+a$).
This latter scenario can be imagined to be rare in most bacteria-phytoplankton interactions, as it requires small phytoplankton cells to produce extremely strong chemoattractant fields. 

\section*{Scaling analysis of the chemotactic performance landscape}
The approximate location of the boundaries of chemotactic detection can be easily obtained in the limit where there is no background chemoattractant concentration ($C_0=0$).
The right boundary (i.e., large $R$), associated with the inherent component of the sensing noise (\autoref{eq:noise_mw}), is identified by the condition that the sensory horizon matches the radius of the phytoplankton cell, $S\sim R$. In this case, \autoref{eq:snr_explicit}, with $C_0=0$ and assuming $f=1$ since the inherent component of the sensing noise does not filter out the high-frequency signals~(\autoref{eq:noise_mw}), reads
\begin{equation}
    \dfrac{1}{\Pi}\dfrac{UT}{R}\sqrt{\dfrac{\pi a D_C T C_S}{3}} = q.
\end{equation}
When solved for $C_S$, this equation gives, as a function of $R$ and all the other parameters, the minimum $C_S$ required to have a sensory horizon at least as large as the cell radius:
\begin{equation}\label{eq:right_boundary}
    \hat C_S^\text{right} = \dfrac{3}{\pi}\dfrac{(q\Pi)^2}{a D_C T}\qty(\dfrac{R}{UT})^2.
\end{equation}
The left boundary (i.e., small $R$), is associated instead with the dynamic component of the sensing noise, and can be identified by the condition that the sensory horizon is limited by the distance that the bacterium travels during a single sensory interval, $S\sim UT$.
In this case, if we take $S\approx UT$ to be the limit where the low-pass filter is just starting to kick in we can again assume $f\approx1$, and, with $C_0=0$, \autoref{eq:snr_explicit} becomes
\begin{equation}
    \dfrac{1}{\Pi}\sqrt{\dfrac{\pi a D_C T C_S}{3}\dfrac{R}{UT}} = q
\end{equation}
from which
\begin{equation}\label{eq:left_boundary}
    \hat C_S^\text{left} = \dfrac{3}{\pi}\dfrac{(q\Pi)^2U}{a D_C R}.
\end{equation}

\autoref{eq:right_boundary} and \autoref{eq:left_boundary} show that the sensitivity threshold, $q$, chemotactic precision, $\Pi$, chemoattractant diffusivity, $D_C$, and bacterial radius, $a$, display the same functional relationships on the left and right boundary. An increase in $a$ or $D_C$ increases the adsorption rate of chemoattractant molecules, therefore decreasing molecule counting noise and allowing bacteria to perform a more robust sampling of the local concentration, and in turn allowing bacteria to detect gradients even when the phytoplankton leakage rate (and therefore $C_S$) is lower. Increasing $\Pi$, which represents additional sources of noise downstream in the signal processing pathway, instead obviously decreases the sensitivity of bacteria to signals. Similarly, higher values of $q$ impose a larger threshold on detection. Since we keep $\Pi=6$ and $q=1$ constant, in the main text we replaced the prefactor $3q^2\Pi^2/\pi \approx 34.4$.
By contrast, we find that the phytoplankton radius, $R$, the bacterial swimming speed, $U$, and the bacterial sensory timescale, $T$, play different roles at the two ends of the size spectrum.
On the right boundary (\autoref{eq:right_boundary}), the minimum $C_S$ required for gradient detection decreases quadratically with speed ($\sim U^{-2}$) and with the third power of the sensory timescale ($\sim T^{-3}$), consistent with the fact that faster motion and longer integration allow the bacteria to sample larger variations in spatially extended concentration field, thus strongly increasing their sensitivity. Increasing the phytoplankton radius $R$ instead increases the minimum $C_S$, consistent with the previous discussion about the size-dependent steepness of the diffusive concentration field.
Surprisingly, on the left boundary (\autoref{eq:left_boundary}), the sensory timescale $T$ appears to not affect the limits of detection at all: the increase in sampling time and the decrease in spatial resolution cancel each other out. However, the value of $T$ does affect the transition from confinement-limited to steepness-limited detection, since the apex of the convex performance landscape is found at $R^*=UT$, and also the values of the $I_C$ within the region of detectability (as shown for example in Figure 4A in the main text).
Moreover, the effect of $U$ and $R$ on sensitivity is inverted: the minimum $C_S$ required for detection increases with $U/R$, so that smaller phytoplankton is harder to detect, and increasing speed makes detection even harder.

\autoref{fig:ic_scalings} shows the scalings of \autoref{eq:right_boundary} and \autoref{eq:left_boundary} overlayed on the chemotactic performance landscape. At $C_0=0$ (\autoref{fig:ic_scalings}A), the scalings perfectly match the boundaries of the landscape, and even with low background concentrations such as $C_0=\SI{1}{\nano M}$ the agreement is good (\autoref{fig:ic_scalings}B). Upon further increasing $C_0$, however, the boundaries are pushed towards higher values of $C_S$ and the simple scalings provide a less precise bound for detection (\autoref{fig:ic_scalings}C). The case with $C_0\neq 0$ cannot be, however, treated in such a simple manner, and \autoref{eq:snr_explicit} must then be solved numerically as we have done in the main text.

\autoref{fig:ic_scalings}D--F show how the position of the boundary moves as a function of $U$, $T$ and $D_C$, consistent with the behavior observed in Figure~4A--C in the main text.

\section*{The Kolmogorov-Smirnov sensor}
In Fig.~4(D--F) in the main text, we have shown numerical estimates of the chemotactic index based on simulations of an idealized sensor moving in a gradient and identifying positive gradients through a Kolmogorov-Smirnov test.

The idealized sensor is defined as a sphere of radius $a$ moving radially towards the source in the concentration field $C(r) = C_0 + C_S\dfrac{R}{r}$. As it moves with a constant speed $U$, the sensor is exposed to the local concentration field and, over a time interval $T$, registers the timeseries of molecular absorption events occurring with instantaneous rate $4\pi D_C a C(r)$.
For any pair of values $(C_S, R)$, the sensor is initialized at a position $r_0 = \max(20R,\SI{100}{\micro\m}) + \delta$ away from the source, where $\delta$ is a random number uniformly distributed over $[-UT/2,+UT/2]$. We introduce this random perturbation to reduce bias in the measurements. If the mesh was fixed, it would be possible to have a transect consistently end up exactly at the source surface, which would always lead to extremely high gradient measurements, even in a scenario where many of the nearby points would give much lower signals. The introduction of the random perturbation decreases the weight associated with some special points in space, providing more realistic outcomes for the measurement process. 

During every measurement interval $T$, the sensor travels the distance $UT$, collecting the time series of absorption events (\autoref{fig:ks}A). To determine whether a positive gradient is present, the sensor compares the cumulative distribution function (CDF) of waiting times in the first half of the interval ($t\in[0,T/2]$), be it $I_1$, with the distribution in the second half of the interval ($t\in[T/2,T]$), be it $I_2$ (\autoref{fig:ks}B).
The comparison is performed through a one-tailed Kolmogorov-Smirnov test, with the following null hypothesis: \textit{the true CDF of the waiting times in the second half of the interval ($I_2$) is not greater than the true CDF in the first half of the interval ($I_1$)}.
In other words, the null hypothesis is that there is no positive concentration gradient.
If we can reject the null hypothesis, then we can conclude that the sensor has detected a positive gradient: the CDF being greater means that the typical waiting times have decreased and therefore the concentration is increasing.

For an individual sensor to reject the null hypothesis, we request a significance level $\alpha=0.05$. The outcome of a single-sensor measurement is therefore a binary value for each interval in the transect, representing whether a gradient was detected in that interval or not (\autoref{fig:ks}C, pink dots). After repeating the transect for an ensemble of $N$ sensors (each one with a different perturbation $\delta$ to its transect), their outcomes are averaged to obtain a curve $f(r)$ which represents the fraction of sensors that detected a gradient around position $r$ (\autoref{fig:ks}C, purple curve). $f(r)$ is a monotonically decreasing function of $r$.
We then define the sensory horizon, $S$, of the Kolmogorov-Smirnov sensor as the farthest distance $r$ from the source at which $f(r) > \bar{f}$, where $\bar{f}$ is an arbitrarily chosen ``consensus threshold'' (\autoref{fig:ks}C, dashed orange line). In other words, $S$ is the largest distance at which at least a fraction $\bar{f}$ of the sensors is able to detect gradient. After obtaining this estimate of $S$, we compute $I_C$ in the same way as for the bacterial model, using \autoref{eq:ic_general}.

In Fig.4D--F, we tune the choice of $\bar{f}$ to get as close as possible to the predictions of the bacterial model based on Equation (3) in the main text. We find that $\bar{f}=0.99$ reproduces a boundary of detection that is rather close to the bacterial model; this is a really high consensus threshold that requires nearly all sensors to agree on the presence of a gradient at a given position.
We find that, without explicitly imposing any form of noise (besides the perturbations to the transect mesh), the performance landscape of the Kolmogorov-Smirnov sensors reproduces the qualitative features we highlighted in Fig.2 in the main text:
for small sources, detection is limited by dynamic noise acting as a low-pass filter, while for large sources, the limiting factor is the inherent noise in the molecular adsorption events, which acts instead as a high-pass filter. The different signal processing performed by this ideal sensor corresponds to a different low-pass filter than the one we assumed for the bacterium in \autoref{eq:snr}; this leads to a slighly different position of the boundary of detection in the low-end of the phytoplankton size spectrum, but does not affect the features of the performance landscape described by our theory (as we show in Figure 4 in the main text).

\section*{Phytoplankton communities: from chemotactic index to search times}
The chemotactic index can be used to obtain an estimate of the average search time experienced by a bacterium in the search for a phytoplankton cell of a given radius $R$, provided that we know the abundance of phytoplankton cells in the environment.
The abundance of phytoplankton cells in marine environments is typically provided as a size spectrum, representing the number of cells within a given size range that will be found in a milliliter of seawater. This means that, in general, we cannot know how many cells having a radius of ``exactly'' $R$ are present in a community, but we can know how many cells have a radius between some finite values $R_1<R<R_2$.
The phytoplankton size spectra usually follow power-law scalings of the form $N(R)\propto R^{-3\alpha}$, where the allometric exponent $\alpha>0$ determines the steepness of the community structure and varies as a function of multiple environmental factors~\cite{maranon2004significance}. Due to the implicit size binning of the phytoplankton radii, $N(R)$ does in fact represent the abundance of cells with sizes in a range $R_1<R<R_2$.
In field observations, the width of these size bins is typically set by experimental limitations; depending on instrumentation used and statistical constraints, the spectra may be composed of $\sim10$ to $\sim30$ bins ranging from a minimum size of $\sim\SI{0.5}{\micro\m}$ up to a maximum size of $\sim\SI{100}{\micro\m}$, with the size of the bins increasing exponentially with the radius of the cells (for details see discussion in Ref.~\cite{haentjens2022phytoplankton}). A finer binning leads to a lower number of cells within each size class, whereas a coarser binning bundles together a larger number of cells within each size class. The choice of the binning therefore also constrains the evaluation of bacterial search times. We consider our phytoplankton size spectra as composed by 17 bins (equally spaced in log-scale), ranging between the two extremal radii of \SI{0.5}{\micro\m} and \SI{70}{\micro\m}, and the total abundance $N_\text{tot}$ is set by multiplying the $R^{-3\alpha}$ scaling by a prefactor in such a way that the sum of the numbers of phytoplankton cells across all the size bins returns $N_\text{tot}$.

Once the size binning has been defined, the average search time experienced by one bacterium for a phytoplankton cell with size $R_1<R<R_2$, with $R$ being the geometric mean of $R_1$ and $R_2$ ($R=\sqrt{R_1R_2}$) can then be evaluated as
\begin{equation}
    T_e = \dfrac{1}{\Gamma N(R)}
\end{equation}
where $\Gamma$ is the kernel for the bacterium-phytoplankton encounter. $\Gamma$, which depends on the phytoplankton radius $R$, is also taken as the geometric mean $\Gamma(R)=\sqrt{\Gamma(R_1)\Gamma(R_2)}$.

\section*{Impact of chemoattractant diffusivity on chemotactic search times}
When considered across the range of potential chemoattractants in the ocean, our analysis reveals an additional surprising result: in certain scenarios, chemoattractants with lower diffusivity cause better chemotactic performance than ones with higher diffusivity (Movie S2). This is surprising because, for a fixed chemoattractant concentration, decreasing $D_C$ decreases the rate at which chemoattractant molecules reach the bacterium, and therefore decreases chemotactic sensitivity (see Equation 4 and Figure 4C in the main text). However, for a fixed leakage rate $L$ of the phytoplankton, a smaller value of $D_C$ implies a higher excess concentration at the phytoplankton surface, $C_S$, because $C_S=L/(4\pi D_CR)$ (\autoref{eq:concentration} and \autoref{eq:leakage}). A higher value of $C_S$, in turn, improves the contrast between the concentration profile around the phytoplankton cell and the background concentration $C_0$. This improved contrast can compensate for the loss in the adsorption rate. We find that this effect is particularly acute for small phytoplankton cells (Movie S2). Slow swimmers may therefore be able to further reduce the search times for the smallest phytoplankton cells in oligotrophic waters if low-diffusivity compounds are exuded at high rates.

We stress that this prediction relies on the assumption of a fixed leakage rate for compounds with different diffusivity, which, however, may not be generally valid. In fact, if we want to interpret a decrease in chemoattractant diffusivity as an increase in the molecular weight of the compound, then the increase in molecular weight should be reflected in an increase in $n_C$, the number of carbon atoms in the compound, which appears in \autoref{eq:leakage_carbon}. This would in turn reduce the leakage rate $L$. The net effect would then depend on the details of the compound structure, i.e. on the relationship between the number of carbons it contains and its diffusivity.
Our understanding of chemotaxis towards high molecular weight compounds such as polysaccharides is still rather limited and warrants further investigation, especially in light of growing evidence of their importance in oceanic carbon sequestration~\cite{buck-wiese2023fucoid} and of their potential to act as strong chemoattractants in the ocean~\cite{clerc2023strong}.

\section*{Average search times may be long, but encounters are plentiful}
The search times we evaluated using Equation~6 in the main text (and reported in Figure~5), represent the average time required by one bacterium to encounter a phytoplankton cell in a given size class.
However, as pointed out in the main text, there are additional factors to consider in order to correctly interpret the meaning of these values, namely (i) assuming a Poisson statistics for the encounter events skews the ``typical'' search time to values lower than the average search time, (ii) the fact that a bacterium is simultaneously exposed to encounters with phytoplankton of all size classes, so that to estimate the total encounters experienced by a bacterium one needs to integrate over the entire size spectrum, and (iii) that even though an individual bacterium may be exposed to relatively few encounters throughout its lifespan, there are at least tens of thousands of other individuals in each milliliter of seawater doing the same, which sum up to plenty of encounters.

Here we break down these factors in order to provide a more complete picture of the encounter landscape in the ocean.

\textbf{Encounter statistics.}
If encounters can be considered as independent events, i.e. there are no spatial correlations between phytoplankton positions in the water column, then the search times are exponentially distributed, i.e., they follow the probability density function
\begin{equation}
    P(t) = \dfrac{1}{T_e}\exp(-\dfrac{t}{T_e})
\end{equation}
where $T_e$ is the average search time, for given bacterial and phytoplankton phenotypes and environmental conditions. The probability that an encounter occurs with a search time shorter than a value $\tau$ is therefore given by
\begin{equation}
    1-\int_0^\tau \dd{t} P(t) = 1 - \exp(-\dfrac{\tau}{T_e}).
\end{equation}
This means that, for a given average search time $T_e$, 50\% of the encounters occur in less than $T_e\ln2\approx0.7T_e$, and 10\% occur in less than $-T_e\ln0.9\approx T_e/10$.

When this statistics is considered, it appears clear that, although typical lifetimes for marine bacteria mostly range from half a day to a few days~\cite{kirchman2016growth}, even encounters with average search times $T_e=\SI{1}{week}$ may be within reach for a sizeable fraction of a bacterial population, where 10\% of the individuals may be able to experience such an encounter in times shorter than \SI{18}{hours}. At the same time, encounters characterized by short average search times, e.g. 10 hours to a day, may be expected to play a particularly prominent role in the dynamics of bacterial population, as even a single bacterium may be able to experience many of them within its lifetime.

The assumption of Poissonian statistics is a convenient one, but it is known to not apply exactly to random walk models of encounters~\cite{hutchinson2007use}, because it ignores the possibility that, after an initial encounter, there may be multiple re-occurrences of encounters due to the now small spatial separation between the bacterium and the phytoplankton. A similar problem has also been considered by Berg and Purcell~\cite{berg1977physics} in their study of molecule counting in bacterial chemotaxis. These corrections can be accounted for by more sophisticate encounter models, but such type of analysis is beyond the scope of the present work.

\textbf{Encounters experienced by an individual bacterium.}
The average encounter rate experienced by a single bacterium can be estimated as the sum of the inverse search times for each phytoplankton size class in the spectrum:
\begin{equation}\label{eq:tot-encounters_single}
    E_1 = \sum_R \dfrac{1}{T_e(R)} = \sum_R \Gamma(R) I_C(R) N(R).
\end{equation}
If, for given bacterial phenotypes, environmental conditions and size structure, we evaluate \autoref{eq:tot-encounters_single} for different values of phytoplankton PER, we can obtain an estimate of how many encounters, on average, we can expect a bacterium to experience each day in its environment.
For a slow swimmer (as in Figure~5, i.e., $U=\SI{25}{\micro\m/\s}$ and $T=\SI{250}{\milli\s}$), we estimate between 1 and 13 encounters per day in oligotrophic waters and between 3 and 14 encounters per day in productive waters. For a fast swimmer ($U=\SI{60}{\micro\m/\s}$ and $T=\SI{100}{\milli\s}$) we estimate between 3 and 12 encounters per day in oligotrophic waters, and between 8 and 21 encounters per day in productive waters.

\textbf{Total encounters at the population scale.}
In order to scale our results from a single bacterium to the level of populations, we need to account for the abundance of motile and nonmotile bacteria in a given environment.
Typical abundances of heterotrophic bacteria in oligotrophic and productive waters can be taken to be $B\sim\SI{1e5}{cells/\milli\l}$ and $B\sim\SI{1e6}{cells/\milli\l}$, respectively~\cite{wigington2016reexamination}. Of these, we can expect a typical motile fraction to be $\sim10\%$~\cite{mitchell1995long}, although higher fractions are also likely depending on conditions.
For nonmotile bacteria, encounter rates can be estimated using the diffusive encounter kernel~(\autoref{eq:kernel_diffusive}), where the ``diffusivity'' is not the effective diffusivity resulting from motility, but the diffusivity associated with the passive Brownian motion. The encounter kernel between a nonmotile bacterium of radius $a$ and a phytoplankton of radius $R$ is therefore~\cite{slomka2023encounter}
\begin{equation}\label{eq:brownian-kernel}
    \Gamma^\text{nonmot}(R) = \dfrac{2k_\text{B}T}{3\eta}\dfrac{(a+R)^2}{aR}
\end{equation}
where $k_\text{B}$ is Boltzmann's constant and $\eta$ is the dynamic viscosity of water. Note that here we also included the Brownian diffusivity of the phytoplankton cells, which was always ignored in previous calculations because it was negligible with respect to the contribution of bacterial swimming.

If $\varphi$ is the motile fraction of bacteria, and $E_1^\text{nonmot}$ is the encounter rate for a single nonmotile bacterium (i.e., the equivalent of \autoref{eq:tot-encounters_single} with the Brownian enconuter kernel of \autoref{eq:brownian-kernel}), then the total population encounter rate is
\begin{equation}\label{eq:tot-encounters_population}
    E = \varphi B E_1 + (1-\varphi) B E_1^\text{nonmot}.
\end{equation}

Assuming, in both cases, a motile fraction $\varphi=10\%$~\cite{mitchell1995long}, we find that a population of slow swimming bacteria may experience between \SI{12000}{} and \SI{130000}{} encounters per day per milliliter of seawater in oligotrophic conditions, and between \SI{340000}{} and \SI{1.5} millions in productive conditions (green circles and bars in \autoref{fig:total-encounters}). For a population of fast swimmers, there would be, each day, between \SI{28000}{} and \SI{120000}{} encounters per milliliter of water in oligotrophic conditions, and between \SI{800000}{} and \SI{2}{} millions in productive conditions (violet squares and bars in \autoref{fig:total-encounters}). In both environments, the contribution of the nonmotile fraction of the population to the total encounters is smaller but still comparable to the lower end of the encounters of the chemotactic population: \SI{5000}{} encounters per day per milliliter in oligotrophic waters, and \SI{120000}{} in productive waters (pink diamonds in \autoref{fig:total-encounters}).

\renewcommand{\thefigure}{S\arabic{figure}}
\setcounter{figure}{0}

\begin{figure*}
    \centering
    \includegraphics[width=0.5\textwidth]{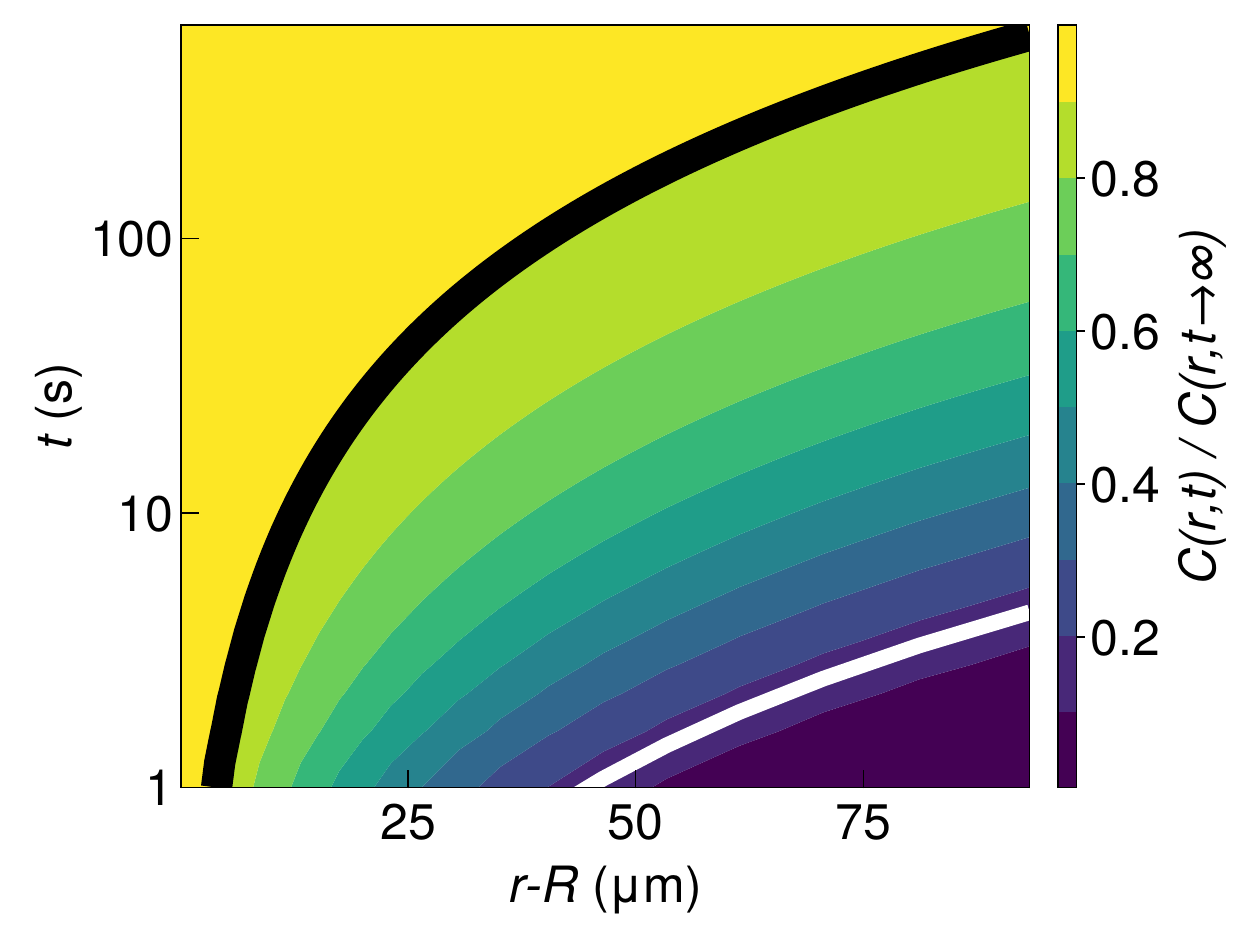}
    \caption{Convergence of the time-dependent solution of the diffusion equation (\autoref{eq:time-dependent-diffusion}) towards the steady state solution (\autoref{eq:concentration}), as a function of distance from the phytoplankton surface ($r-R$) and time.
    The white line is the characteristic timescale $\tau(r)$ (\autoref{eq:diffusive-timescale}) as a function of the distance from the phytoplankton surface, the black line is the locus of points at which the time-dependent solution reaches 90\% of the steady state value, corresponding to approximately $125\tau(r)$.
    }
    \label{fig:time-dependent-diffusion}
\end{figure*}

\clearpage
\begin{figure*}
    \centering
    \includegraphics[width=0.5\textwidth]{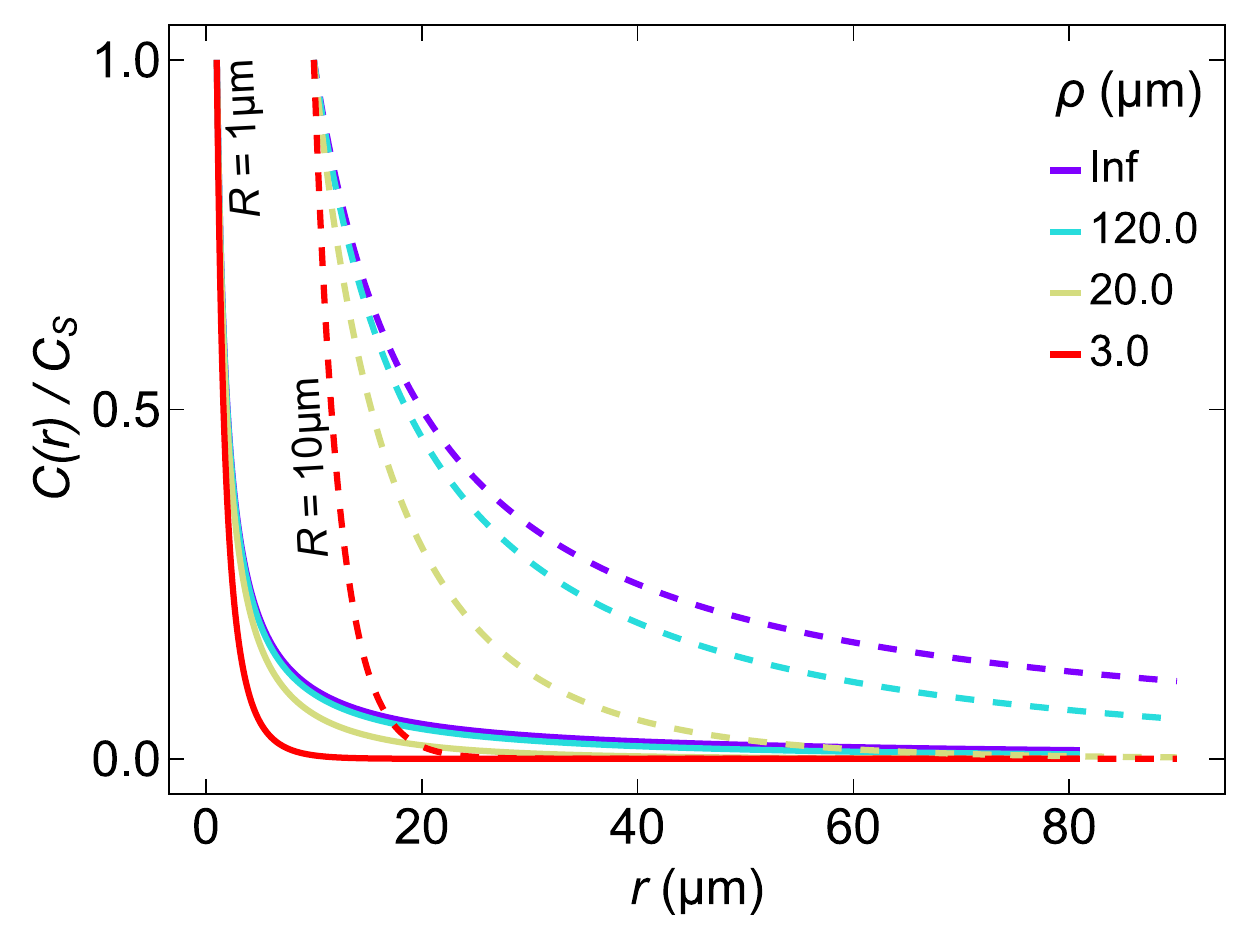}
    \caption{Screened concentration profile (\autoref{eq:expfield}) for different values of $\rho$ (color-coded) and two values of phytoplankton radius ($R=\SI{1}{\micro\m}$, solid, and $R=\SI{10}{\micro\m}$, dashed).
    The $\rho\to\infty$ limit (violet curves) corresponds to the ideal diffusive profile without screening (\autoref{eq:concentration}).
    }
    \label{fig:concentration-fields}
\end{figure*}

\clearpage
\begin{figure*}
    \centering
    \includegraphics[width=\textwidth]{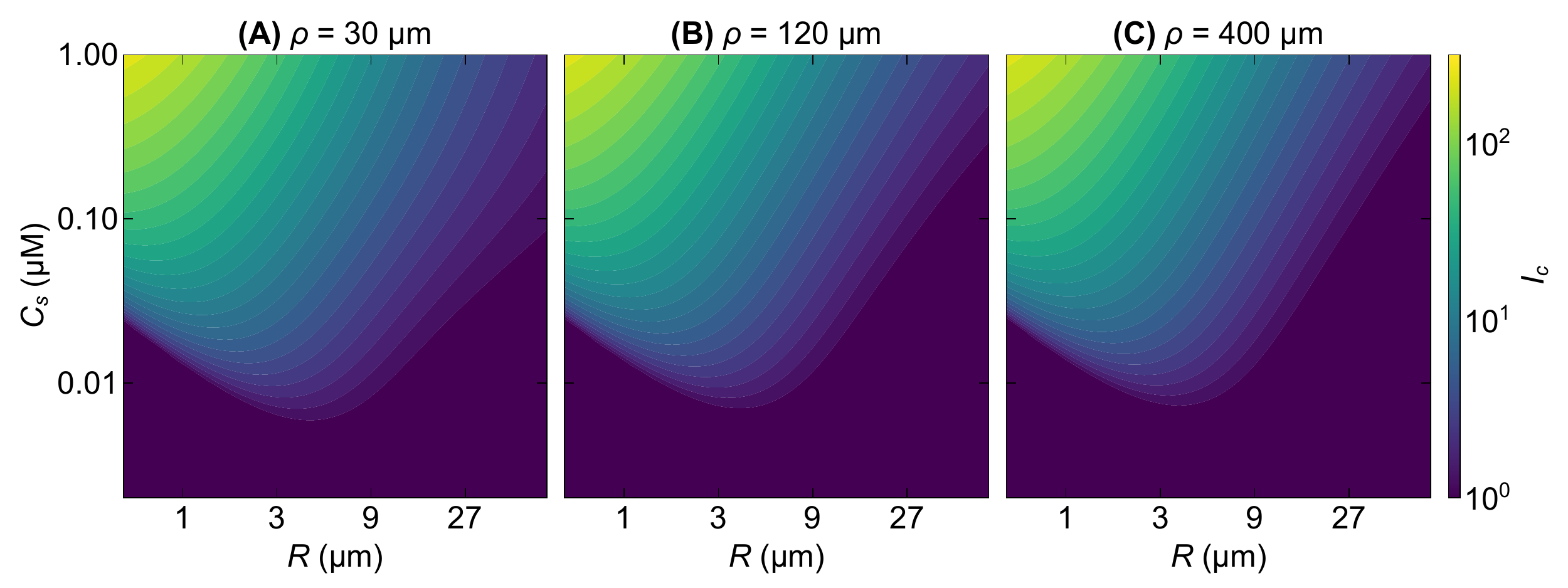}
    \caption{Chemotactic performance landscape for the exponentially decaying concentration field (\autoref{eq:expfield}) for different values of the screening length $\rho$.
    (A) $\rho=\SI{30}{\micro\m}$ corresponds to a bacterial concentration $B\sim\SI{1.8e8}{cells/\milli\liter}$; (B) $\rho=\SI{120}{\micro\m}$ corresponds to $B\sim\SI{1.1e7}{cells/\milli\liter}$ and (C) $\rho=\SI{400}{\micro\m}$ to $B\sim\SI{1e6}{cells/\milli\liter}$.
    The landscape was evaluated for a bacterium with speed $U=\SI{50}{\micro\m/\s}$, radius $a=\SI{0.5}{\micro\m}$, sensory timescale $T=\SI{100}{\milli\s}$, correlation length $\lambda=\SI{30}{\micro\m}$, chemotactic precision $\Pi=6$, a chemoattractant with diffusivity $D_C=\SI{500}{\micro\m^2/\s}$, and background concentration $C_0=\SI{1}{\nano M}$.
    }
    \label{fig:ic_expfield}
\end{figure*}

\clearpage
\begin{figure*}
    \centering
    \includegraphics[width=\textwidth]{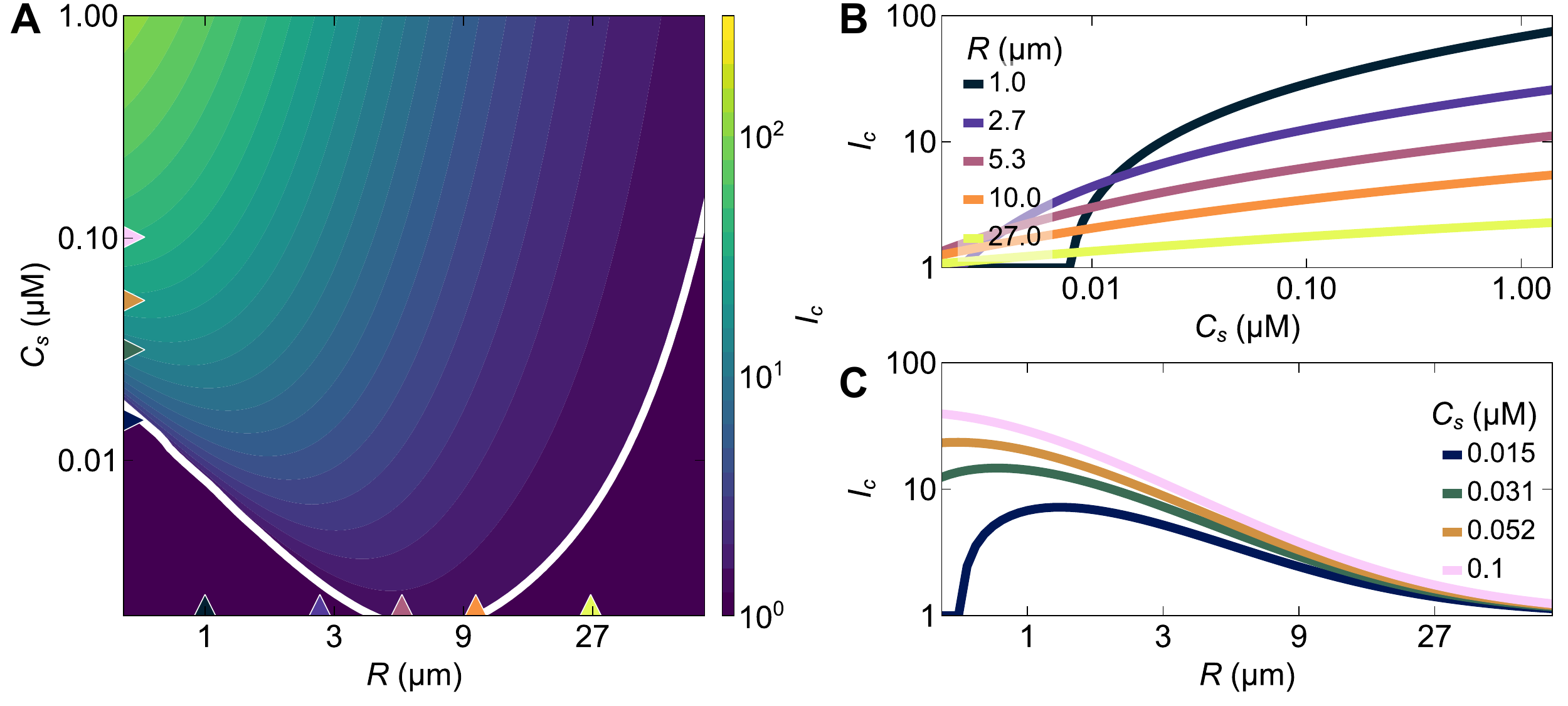}
    \caption{(A) Chemotactic performance landscape in exponentially screened concentration profile (\autoref{eq:expfield}) with screening length $\rho=\SI{3}{\micro\m}$, evaluated for a bacterium with speed $U=\SI{50}{\micro\m/\s}$, radius $a=\SI{0.5}{\micro\m}$, sensory timescale $T=\SI{100}{\milli\s}$, correlation length $\lambda=\SI{30}{\micro\m}$, chemotactic precision $\Pi=6$, a chemoattractant with diffusivity $D_C=\SI{500}{\micro\m^2/\s}$, and background concentration $C_0=\SI{1}{\nano M}$. (B) Vertical transects from panel A for fixed values of the phytoplankton radius $R$. (C) Horizontal transects from panel A for fixed values of the excess concentration $C_S$.
    }
    \label{fig:extreme-screening}
\end{figure*}

\clearpage
\begin{figure*}
    \centering
    \includegraphics[width=\textwidth]{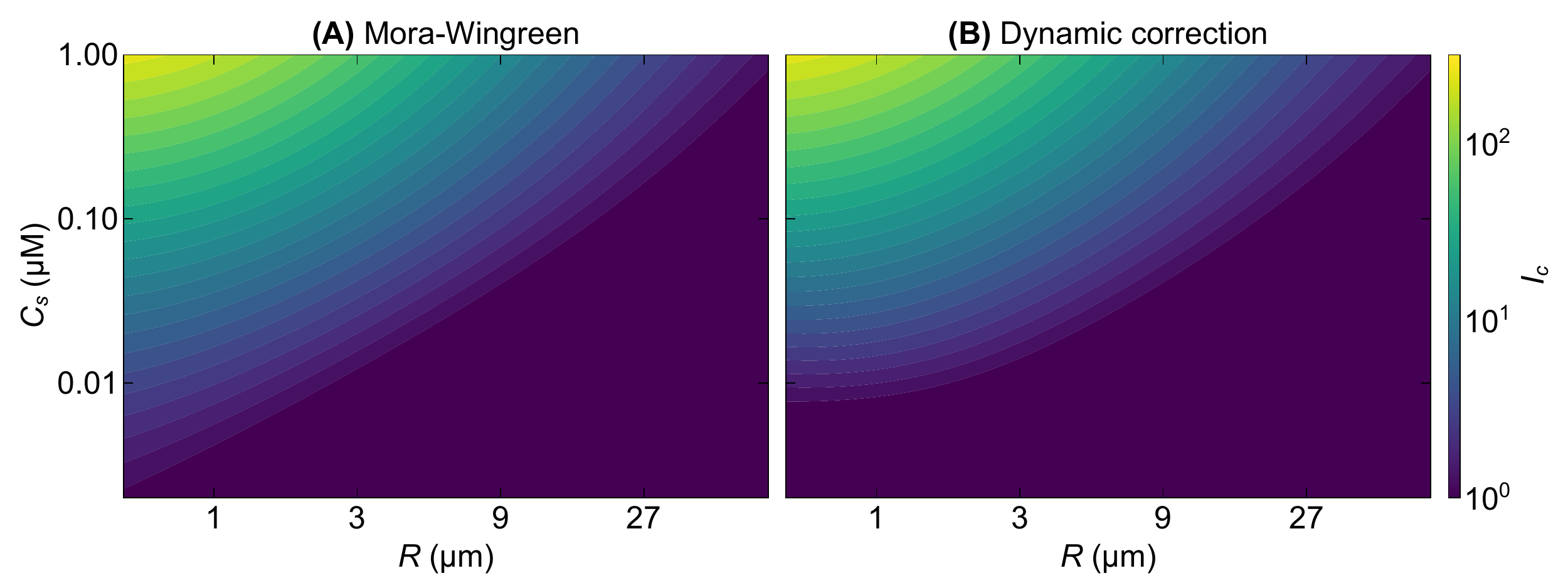}
    \caption{Comparison of the chemotactic performance landscape with and without the dynamic correction to the sensing noise (\autoref{eq:noise_correction}).
    (A) Chemotactic index estimated by only considering the inherent component of the sensing noise evaluated by Mora and Wingree (\autoref{eq:noise_mw}) does not account for the signal degradation due to a finite signal integration window for small phytoplankton cells.
    (B) Chemotactic index estimated by including the dynamic correction to the sensing noise (\autoref{eq:noise_correction}).  The modified landscape shows the increasing role of the signal degradation for small phytoplankton cells, as indicated by the reduction in the size of the region of detection for small $R$, but the dynamic correction (\autoref{eq:noise_correction}) does not fully capture the rate at which the signal is degraded for the smallest cells.
    The landscape was evaluated for a bacterium with speed $U=\SI{50}{\micro\m/\s}$, radius $a=\SI{0.5}{\micro\m}$, sensory timescale $T=\SI{100}{\milli\s}$, correlation length $\lambda=\SI{30}{\micro\m}$ and chemotactic precision $\Pi=6$, and a chemoattractant with diffusivity $D_C=\SI{500}{\micro\m^2/\s}$ and background concentration $C_0=\SI{100}{\nano M}$.
    }
    \label{fig:ic_noise}
\end{figure*}

\clearpage
\begin{figure*}
    \centering
    \includegraphics[width=\textwidth]{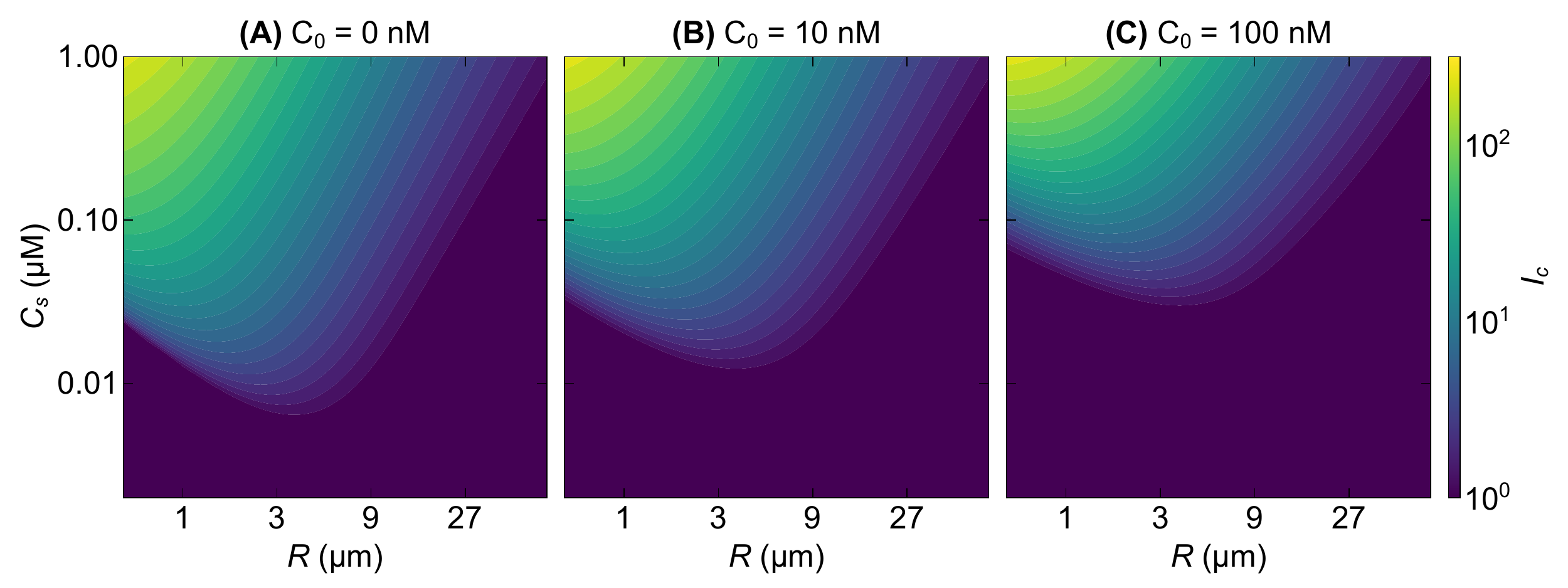}
    \caption{Effect of varying the background cheomattractant concentration $C_0$ on the chemotactic performance landscape.
    Larger values of $C_0$ are associated with larger sensing noise (\autoref{eq:noise_mw}); the boundary of detection is therefore shifted towards larger $C_S$ values because stronger gradients are required to overcome the additional noise. The convex shape and the asymmetry of the chemotactic performance are, however, not affected by the value of $C_0$.
    The landscape was evaluated for a bacterium with speed $U=\SI{50}{\micro\m/\s}$, radius $a=\SI{0.5}{\micro\m}$, sensory timescale $T=\SI{100}{\milli\s}$, correlation length $\lambda=\SI{30}{\micro\m}$, chemotactic precision $\Pi=6$, and a chemoattractant with diffusivity $D_C=\SI{500}{\micro\m^2/\s}$.}
    \label{fig:ic_c0}
\end{figure*}

\clearpage
\begin{figure*}
    \centering
    \includegraphics[width=\textwidth]{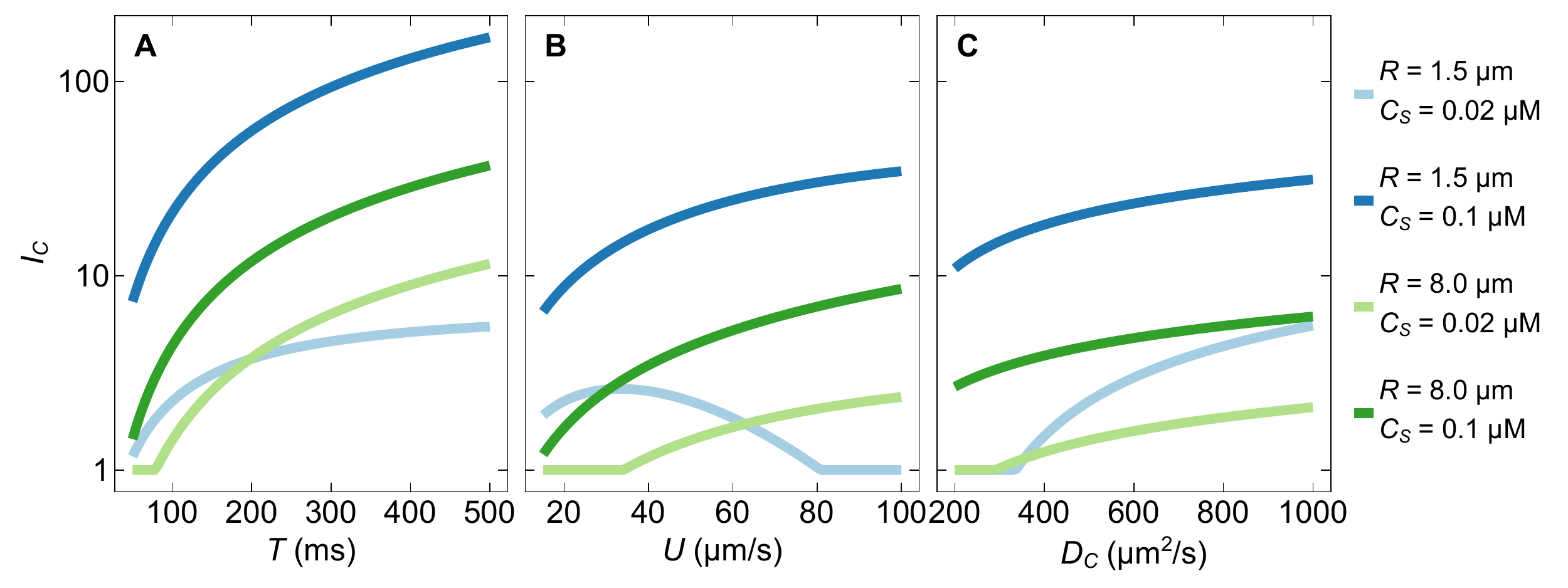}
    \caption{Chemotactic index as a function of (A) bacterial sensory timescale, (B) bacterial swimming speed and (C) chemoattractant diffusivity, for a small and a large phytoplankton cell ($R=\SI{1.5}{\micro\m}$ and $R=\SI{8}{\micro\m}$) exuding chemoattractant with low and high excess concentration ($C_S=\SI{0.02}{\micro M}$ and $C_S=\SI{0.1}{\micro M}$).
    (A) Increasing $T$ always increases the $I_C$, but for the small phytoplankton with low $C_S$ (light blue curve) the increase is less steep and nearly saturates.
    Calculations performed with $U=\SI{50}{\micro\m/\s}$ and $D_C=\SI{500}{\micro\m^2/\s}$ (as in Figure 3 of main text).
    (B) When $C_S$ is large, increasing $U$ increases the $I_C$ for both the small and large phytoplankton (dark blue and dark green curves). When $C_S$ is low, however, the large phytoplankton (light green curve) can only be detected ($I_C>1$) with a sufficiently large $U$ ($\approx\SI{35}{\micro\m/\s}$), whereas the small phytoplankton (light blue curve) can only be detected for $U$ lower than $\approx\SI{80}{\micro\m/\s}$ and the $I_C$ shows a maximum at $U\approx\SI{25}{\micro\m/\s}$.
    Calculations performed with $T=\SI{100}{\milli\s}$ and $D_C=\SI{500}{\micro\m^2/\s}$.
    (C) Increasing $D_C$ always increases the $I_C$ but the effect is most noticeable for the small phytoplankton with low excess concentration (light blue curve): for low $D_C$ the phytoplankton cannot be detected but the $I_C$ sharply increases when $D_C>\SI{350}{\micro\m^2/\s}$.
    Calculations performed with $T=\SI{100}{\milli\s}$ and $U=\SI{50}{\micro\m/\s}$.
    In all panels, we fix radius $a=\SI{0.5}{\micro\m}$, correlation length $\lambda=\SI{30}{\micro\m}$, chemotactic precision $\Pi=6$, and background concentration $C_0=\SI{10}{\nano M}$.}
    \label{fig:ic_TUD}
\end{figure*}

\clearpage
\begin{figure*}
    \centering
    \includegraphics[width=\textwidth]{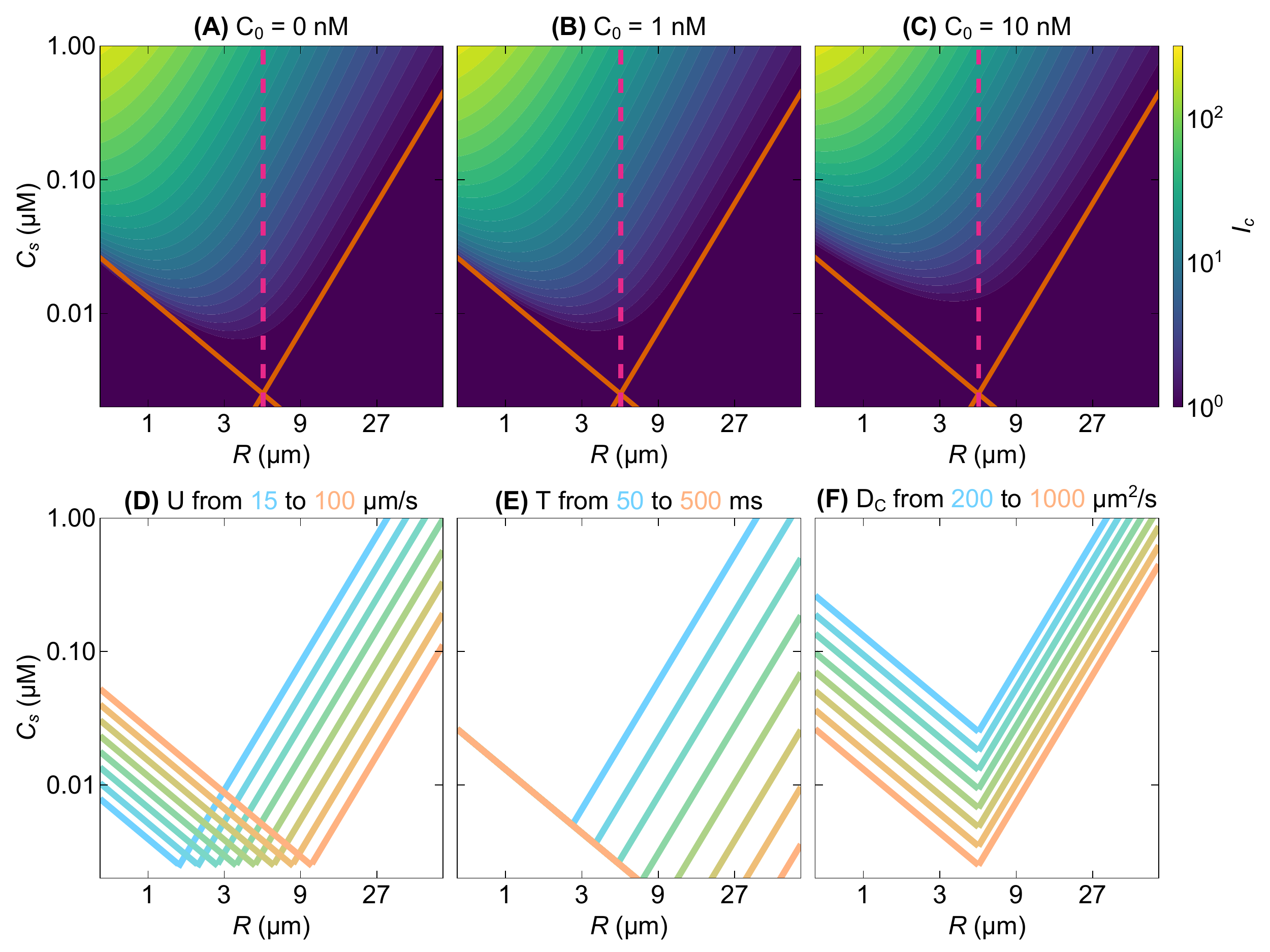}
    \caption{Scaling calculation of the position of the detection boundary. (A-C) The boundary positions are superimposed on the chemotactic performance landscape for different values of the background chemoattractant concentration $C_0$.
    The solid orange lines represent the boundary locations evaluated under the assumption that $C_0=0$ (\autoref{eq:right_boundary} and \autoref{eq:left_boundary}). The dashed magenta line denotes the point where the two lines intersect, $R^*=UT$, corresponding to the phytoplankton radius for which the excess concentration $C_S$ required for gradient detection is minimal.
    The landscape was evaluated for a bacterium with speed $U=\SI{50}{\micro\m/\s}$, radius $a=\SI{0.5}{\micro\m}$, sensory timescale $T=\SI{100}{\milli\s}$, correlation length $\lambda=\SI{30}{\micro\m}$, chemotactic precision $\Pi=6$, and a chemoattractant with diffusivity $D_C=\SI{500}{\micro\m^2/\s}$ and using a sensitivity threshold $q=1$ on the $\SNR$.
    (D-F) Variation of the boundary position upon changing a single parameter at a time, with values equally spaced on a logarithmic scale. In all three panels, the colors range from cyan for the lowest value to orange for the highest value.}
    \label{fig:ic_scalings}
\end{figure*}

\clearpage
\begin{figure*}
    \centering
    \includegraphics[width=\textwidth]{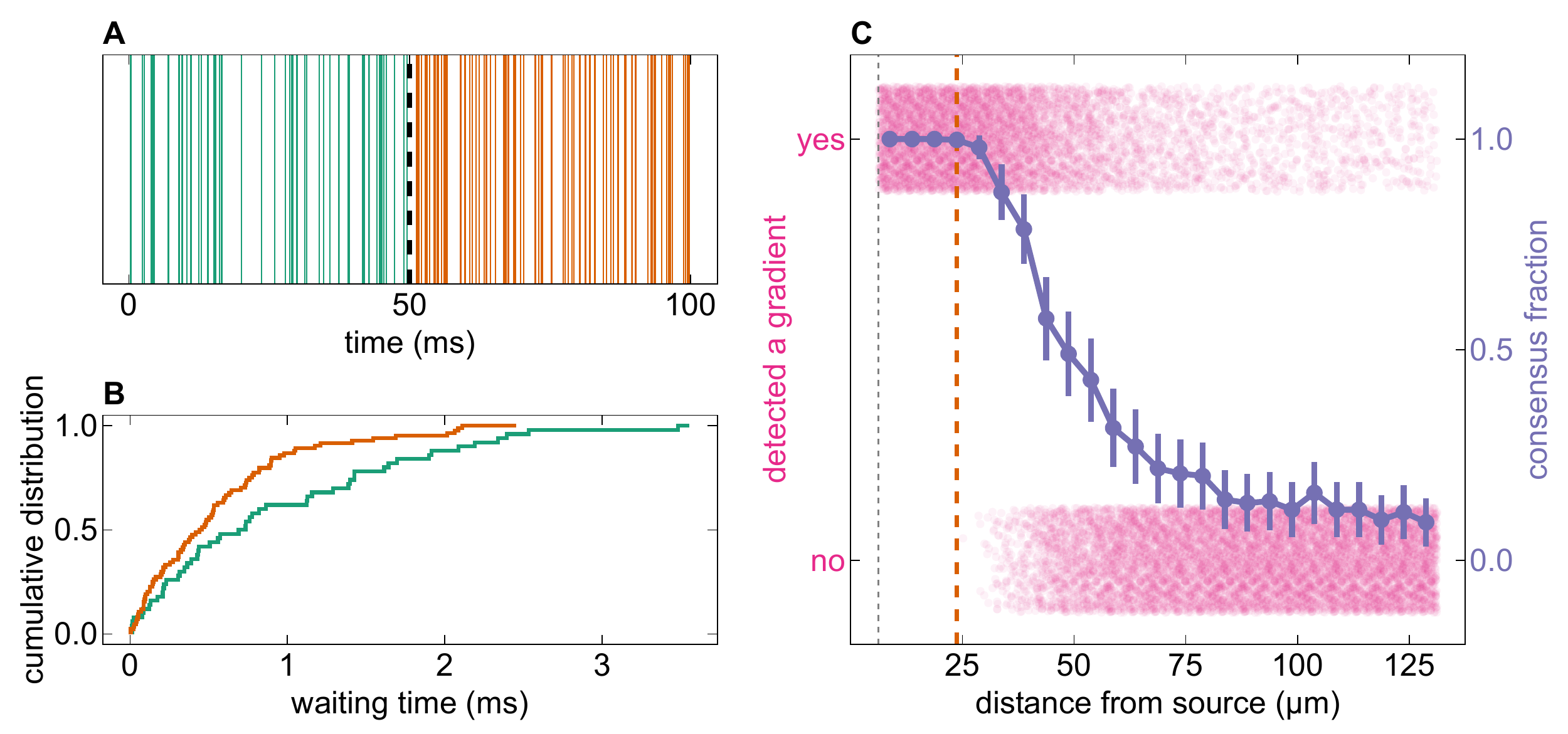}
    \caption{Evaluation of the sensory horizon via the Kolmogorov-Smirnov model sensor.
    (A) Timeseries of adsorption events experienced by a sensor moving in a concentration field over a sensory interval $T=\SI{100}{\milli\s}$. Events that happened in the first and second half of the sensory interval are highlighted in green and orange, respectively.
    (B) Cumulative distribution of the waiting times experienced by the sensor in panel A.
    The cumulative distributions (whose colors match the timeseries in panel A) are subjected to the Kolmogorov-Smirnov test to assess if, at the current location, a chemical gradient is present.
    (C) Outcomes from an ensemble of 500 sensors performing a transect towards a source.
    Each pink dot represents the binary outcome of a single 'yes/no' gradient measurement event at different positions in space (left axis). That is, each point is obtained through the process schematized in panels A and B as the sensor moves through space. The dots are slightly displaced along the vertical axis for visualization purposes. 
    The solid purple line is the consensus fraction (right axis) in an ensemble of sensors (i.e., the fraction of sensors that detected a gradient at a given position), and the error bars are the associated standard errors.
    The thin grey dashed line is the radius of the source ($R=\SI{6.3}{\micro\m}$ in this example), and the thick orange dashed line is the sensory horizon (here $S\simeq\SI{24}{\micro\m}$) estimated as the distance from the source where the consensus fraction becomes larger than a threshold of 0.99.
    }
    \label{fig:ks}
\end{figure*}

\clearpage
\begin{figure*}
    \centering
    \includegraphics[width=0.5\textwidth]{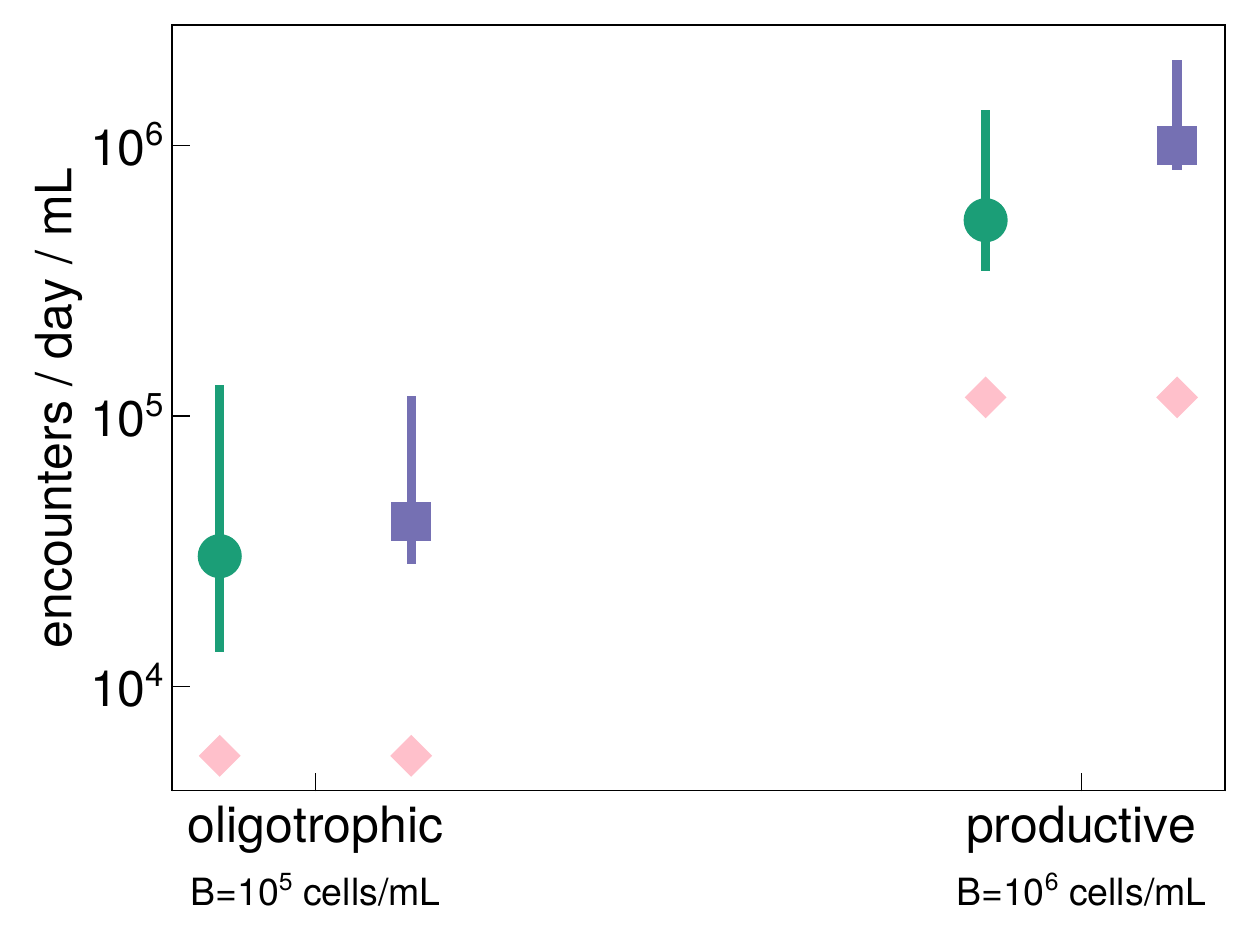}
    \caption{Total bacteria-phytoplankton encounters occurring per day in a milliliter of seawater, estimated according to \autoref{eq:tot-encounters_population}.
    Calculations are performed for oligotrophic and productive environments (identical to those used in Figure~5 in the main text), with total bacterial abundances $B=\SI{1e5}{cells/\milli\l}$ and $B=\SI{1e6}{cells/\milli\l}$, respectively. In both cases, it is assumed that only a fraction $\varphi=0.1$ of the bacterial population is motile. The pink diamonds represent the encounters of nonmotile bacteria (\autoref{eq:brownian-kernel}), while the green circles and violet squares represent the encounters of slow swimmers and fast swimmers, respectively (same as Figure~5) with a phytoplankton PER of 10\%. The bars represent the range associated with PER values from 2\% to 40\%.
    }
    \label{fig:total-encounters}
\end{figure*}

\clearpage
\bibliography{library_rf}
\bibliographystyle{apsrev4-1}

\end{document}